\documentstyle[amssymb,12pt,epsfig]{article}
\title {The $\nu_\mu \leftrightarrow \nu_s$  interpretation
of   the atmospheric  neutrino data  and
cosmological constraints.}

\author{Pasquale Di Bari, Paolo Lipari, Maurizio Lusignoli \\
I.N.F.N., Sezione di Roma, and \\
Dipartimento di Fisica, Universit\`a di Roma ``la Sapienza",\\
P. A. Moro 2,  I-00185 Roma, Italy}

\begin{document}

\maketitle

\begin{abstract}
The  data on atmospheric  neutrinos  can  be  explained
assuming the existence  of oscillations 
between  $\nu_\mu$'s and  a   light sterile
neutrino  with   mixing close  to  maximal, and  
$\delta m^2 \sim 3 \times 10^{-3}$~eV$^2$.
This interpretation  of the data is  in potential  conflict
with  the successes of  big bang nucleosynthesis (BBN),
since oscillations can result in  a too large  contribution 
of the sterile  state to the energy density of the  universe at the  epoch of 
nucleosynthesis.
The possibility to evade these   cosmological    constraints
has been  recently  the object of some controversy.
In this  work  we   rediscuss this problem
and  find that  the  inclusion  of  a small  mixing
of the sterile  state  with $\nu_\tau$    can result
in    the  generation of  a     large
lepton asymmetry   that  strongly  suppress the
$\nu_\mu \leftrightarrow \nu_s$ oscillations
eliminating the possible conflict with BBN bounds.
In this  scheme  the mass  of the tau neutrino  must be  
larger than few eV's    and is  compatible  with  cosmological  bounds.
Our  calculations  is  performed  using  a  Pauli-Boltzmann  method.
In this  approach  it is also possible   to develop  analytic  calculations
that  allow physical   insight in  the processes  considered  and
give support to the numerical  results.
\end{abstract}

\section{Introduction}
The  data  on  atmospheric neutrinos  \cite{SK,atm-other}
have   shown  that   muon  neutrinos and  antineutrinos
`disappear'    oscillating
with a   large mixing   parameter    $\sin^2 2 \theta \simeq 1$
and  a squared mass difference $|\delta m^2| \simeq   
3  \times 10^{-3}$~eV$^2$ 
into  an `invisible' state.   Experimentally  
this  state can   be  either a  $\tau$ or  a sterile  neutrino.
The   existence of  light  sterile  neutrino states  is also 
suggested  by the results on solar  neutrinos \cite{solar}
and of the LSND experiment \cite{LSND}  that  give  independent   hints for
neutrino oscillations.

If the observed  oscillations  are into  a  sterile state, 
this  could   have   significant  cosmological  implications,
and  in fact  it has   been  argued  
\cite{Barbieri90,Enqvist92b,Shi93} 
that  oscillations  with   these
parameters  are  already excluded   by 
the present  bounds  obtained  from  primordial nucleosynthesis.
Sterile neutrinos   not   mixed  with ordinary  neutrinos
would    have a small   average density    for  temperature 
$T \lesssim  100$~MeV,  and    therefore have  a negligible  influence
during the epoch of  nucleosynthesis.
However   oscillations   between  sterile  and   standard  neutrinos
can be  a  source of
sterile  particles,   and  indeed   if the oscillation
have sufficiently large  amplitude  and are sufficiently  rapid,
the  sterile states  can  be brought in thermal  equilibrium with
ordinary matter  and    give  a significant contribution to the
energy density.  
The success of  Big Bang Nucleosynthesis (BBN) 
in predicting  the  relative   abudances of 
primordial  nuclear  species 
can be used  to  put   limits   on the   energy density
during the epoch of  nucleosynthesis 
or  on the  effective  number of  light  particles
present  during this  epoch.
For  example recently  Olive, Steigman  and Walker \cite{BBN}
quote a (two sigma) limit  $N_\nu^{\rm eff} \le 3.3$, assuming the so-called 
low--deuterium option. 
Such  a limit  can   be translated   into  an allowed  and 
a forbidden  region   for the  
parameters $\sin^2 2 \theta$ and  $\delta m^2$ 
that describe  the oscillations  of  a sterile  neutrino   with an
ordinary one.
Presently   there is  some  controversy on the  value of the limit
for  $N_\nu^{\rm eff}$;
if the limit is  relaxed  to $N_\nu^{\rm eff} \le 4$, as someone
conservatively supports \cite{BBN_2}, obviously 
no  region in  the neutrino  oscillation parameter  space can 
be excluded,    however it appears  possible 
that a   sterile  neutrino  in   thermal  equilibrium 
with ordinary matter   could be   excluded   by BBN, once the 
controversial datum on primordial deuterium abundance from high
redshift quasar absorbers will be definitely settled \cite{Tytler}.

Assume  now that $N_\nu^{\rm eff} \gtrsim 4$ is   excluded
by the considerations on nucleosynthesis:  does this  also
exclude the   interpretation of
the atmospheric  neutrino  data in terms  of
$\nu_\mu \leftrightarrow  \nu_s$   oscillations ?
We  will  argue here that this is  not the case
providing an explicit  simple model where the 
cosmological  bound is respected.

In this  work    we  will    
assume   that a  sterile neutrino state  is  mixed   with  
the $\nu_\mu$ 
with oscillation parameters   compatible with the fits
to the atmospheric  neutrino data, and we will compute the
contribution to  the  energy density   at the  time  of nucleosynthesis
due to $\nu_s$   produced  by  oscillations.
As  already found  by several authors 
\cite{Barbieri90,Enqvist92b,Shi93},  
if the   sterile  state  is  only
mixed  with   the   $\nu_\mu$,     for the  oscillation   parameters
suggested by the  atmospheric $\nu$  data, 
to a good  approximation, it is  indeed put in   thermal   equilibrium
with the rest of the plasma, 
and the   effective  number  of  neutrinos
during  nucleosynthesis   becomes $N_\nu^{\rm eff} \simeq  4$, in possible 
conflict with the data on primordial  nuclear  abundances.
It is  however remarkable  that the   density   of the sterile  state
depends  critically  on the   details
of its mixing  with other   ordinary  neutrinos.
In particular     we will show  that the
inclusion of  a   very small    mixing of the sterile  state
with  the $\tau$  neutrino
($\sin^2 2 \theta_{\tau s} \simeq 10^{-5}$--$10^{-9}$)
can   reduce   dramatically the  energy density  of the sterile  state
to  a   neglibly  small   value  if  the   neutrino that
is  prevalently  $\nu_\tau$  has a mass $m_{\nu} \gtrsim 1$~eV.
This  effect  has  been   demonstrated   for the   first  time
by Foot and Volkas \cite{Foot97}.
Qualitatively  the  mechanism  that is operating is that 
$\nu_\tau \leftrightarrow \nu_s$ oscillations develop  earlier 
in time  and generate a  net   $\tau$--lepton   number  in the 
plasma   that acts  to  
block  the $\nu_\mu \leftrightarrow \nu_s$ oscillations.

This  work  is organized  as  follows.
In section 2  we describe the    effective  potential  for 
neutrinos  in the hot and  dense  medium  of the early  universe
and    discuss how the oscillation parameters   
for  mixing between  standard and  ordinary  neutrinos 
are  modified  by  the presence of  matter.
In section 3  we discuss    the method  we have used to 
compute the    evolution   with time  of the  populations
of  standard  and sterile  neutrinos.
We have chosen in this  work    not to use the  full  machinery of the 
quantum kinetic equations   but a  simpler  
(Pauli--Boltzmann)  approach,    that is  approximately valid  when  the  
standard neutrino    mean free  path  is 
much  shorter than the oscillation  length.
In section 4  we will  discuss the  evolution  of the sterile  neutrino density
in the presence of   a  two--flavor   mixing  with a standard neutrino, 
considering the    lepton  number  per  unit comoving  volume
as  constant.
In section 5 we   rediscuss the  two--flavor  mixing  problem  
taking into account the  fact  that 
since  neutrinos   and  antineutrinos  in matter  can
oscillate   with different  amplitudes, the oscillations  can  generate 
a  large lepton  number   in the  medium, 
that    acts back on the oscillations  
modifying the effective oscillation  parameters.
In section 6 we discuss the flavor evolution  of  a system  of 
two  standard ($\nu_\mu$ and $\nu_\tau$) and one  sterile neutrino  states,
and  determine  the  region of  oscillation parameters  that
is in  agreement  with   the 
$\nu_\mu \leftrightarrow  \nu_s$  interpretation of the atmospheric  neutrino
data, and  also 
results  in a  small contribution of sterile  neutrinos and
anti--neutrinos  to the energy density of the  universe during nucleosynthesis.

\section{Effective neutrino mixing in the early universe}

We will study the cosmological consequences of assuming the existence
of a two-state mixing between two different flavour neutrinos, $\nu_{\alpha}$ 
and $\nu_{\beta}$, with a vacuum mixing angle $\theta_{0}$
(defined  in the  interval $0 \le  \theta_0 \le \pi/4$)
and a squared mass  difference  $\delta m^2 = m_b^2 -m_a^2$.
This means that the   flavor  eigenstates 
are a linear combination of the mass eigenstates
$\nu_{a}$, $\nu_{b}$  (with well  defined  masses $m_a$  and  $m_b$)
\begin{equation}
|\nu_{\alpha}\rangle=
 \cos\theta_{0}|\nu_{a}\rangle+\sin\theta_{0}|\nu_{b}\rangle,
\hspace{10 mm}
|\nu_{\beta}\rangle=\cos\theta_{0}|\nu_{b}\rangle
-\sin\theta_{0}|\nu_{a}\rangle.
\end{equation}
In this way,
for ultrarelativistic  neutrinos of  momentum $p$ propagating in vacuum,
 the probability that a state $|\nu(t)\rangle$ being a pure 
$|\nu_{\alpha}\rangle$ at an initial time $t_{in}$, is measured as a 
$|\nu_{\beta}\rangle$ at a time $t_{in}+\tau$ is:
\begin{equation}
P_{0}(\nu_{\alpha}\to \nu_{\beta},t_{in}+\tau)\equiv 
|\langle\nu_{\beta}|\nu(t_{in}+\tau)\rangle|^{2}=
\sin^{2}2\theta_{0}\cdot\sin^{2}
\left(\frac{\tau}{2\ell_{0}(p)}\right)
\end{equation}
where we defined the vacuum oscillation length as 
$\ell_{0}(p)\equiv 2p/\delta m^{2}$.
If the neutrinos   propagate in matter,
 the effect of coherent forward scattering of neutrinos with the 
particles of the medium results in a flavor dependent  effective potential
that  modifies  the mixing angle and the oscillation
length 
\cite{Wolf78}:
\begin{equation}
\label{eq:sm}
\sin^{2}2\theta_{m}(p)=\frac{s^{2}}{s^{2}+(c-v_{\alpha}(p)+v_{\beta}(p))^{2}}
\end{equation}
\begin{equation}
\label{eq:lm}
\ell_{m}(p)=\frac{\ell_{0}(p)}{\sqrt{s^{2}+(c-v_{\alpha}(p)+v_{\beta}(p))^{2}}}
\end{equation}
where we introduced the adimensional effective potentials 
$v_{\alpha,\beta}(p)\equiv (2p/\delta m^{2})V_{\alpha,\beta}$ and we defined
$s\equiv\sin 2\theta_{0}$, $c\equiv\cos 2\theta_{0}$. 
Consequently, for  neutrinos  propagating 
in a  homogeneous  medium of  constant  
density and  composition one has that:
\begin{equation}\label{eq:probm}
P_{0}(\nu_{\alpha}\to \nu_{\beta},t_{in}+\tau)\longrightarrow
P_{m}(\nu_{\alpha}\to \nu_{\beta},t_{in}+\tau)=
\sin^{2}2\theta_{m}\cdot \sin^{2}\left(\frac{\tau}{2\ell_{m}(p)}\right) 
\end{equation}	
If the condition $c=v_{\alpha}(p)-v_{\beta}(p)$ is satisfied, then a resonance 
occurs and mixing is maximal \cite{MS86}. The resonant amplification of the 
oscillations was first studied in the solar environment were oscillations 
develop coherently and MSW conversion takes place. 
In the  following we  will  study 
 the mixing of a  standard    neutrino 
($\nu_\alpha$  with $\alpha = e,\mu,\tau)$
 and  a strictly 
sterile  neutrino, 
$v_{\beta}= v_{s} = 0$ in the previous expressions. 
The effective potential of a standard neutrino depends
on its momentum and on the temperature and composition of the medium.

We will  study the effect of  neutrino oscillations during the 
time  when the  universe   has  a  temperature 
$T \lesssim  150$~MeV. This is the range of temperatures   in  which 
the oscillations can have effects on the products of BBN. 
The    description of  neutrino oscillations for higher  temperatures 
and  during the  quark-hadron phase transition is  a difficult  problem,
in any case  the oscillations  at  high temperature  are 
suppressed  by matter  effects   as  it will be discussed in  the following.
Note also  that 
the population of    sterile   neutrinos  produced by
oscillations    at high temperature
would be afterwards diluted by photon production
due to the  decrease in the number of  degrees of freedom. 

We  will assume   for  simplicity 
that the  comoving  density of  photons  is  constant,
and that  the   density of muons  in the medium is  negligible.
These  assumptions  are  not  
exactly correct  in the range of   temperatures we  are  considering,
but are  good  approximations.
The  ratio  of the comoving density of  muons   and
electrons  $N_\mu(T)/N_e(T)$ is  0.8 for  $T = m_\mu \simeq  105$~MeV, and
drops to 0.1 for  $T \simeq 26$~MeV, as the  entropy of the  muons
is  transferred  to the  other particles   of the medium,
including the photons.
Considering that the statistical weight 
of muons is about $20\%$ when they are fully present, we can expect that the 
accuracy of our analysis is $0.02$--$0.2$ for $25 \lesssim T/$MeV~$
\lesssim 150$.
 On the other hand for $T>0.5$~MeV~$\simeq m_{e}$ 
the variation in the electron particle number 
can be neglected with a much better accuracy.

The effective potential for  a  standard neutrino 
of  flavor $\alpha$
can be  written \cite{Raffelt88,Enqvist91}
as the sum  of two terms
\begin{eqnarray}
v_{\alpha}(p,T,L^{(\alpha)}) & = & v_{\alpha}^{L}(p,T,L^{(\alpha)})
		            +v_{\alpha}^{T}(p,T)
\\
& = & {a_L \;T^4 \;y  \over \delta m^2}~ \;L^{(\alpha)}
- {b_\alpha \;T^6 \;y^2 \over \delta m^2}  
\label{eq:vv}
\end{eqnarray}
The effective potential  for  antineutrinos 
is  obtained  changing the sign  of the $v^L$   term,
while  the $v^T$  term  remains  unchanged. 
Note  that in a  medium with  a  net  charge  number  $L^{(\alpha)}$ 
neutrinos  and
anti--neutrinos have different oscillation properties.
In  equation (\ref{eq:vv})
we have defined  $y = p/T$, and  $L^{(\alpha)}$   is a linear
combination  of the  charge asymmetries  (per  photon) of the medium
that  is  relevant for   the oscillations   of   standard
neutrinos of  flavor $\alpha$.
For  muon  neutrinos  we  have 
\begin{equation}
L^{(\mu)} = 
2\, L_{\nu_{\mu}}+L_{\nu_{e}} +L_{\nu_{\tau}}-\frac{1}{2}B_{n},
\label{eq:vlal}
\end{equation}
where  $Q_x = (n_x - n_{\overline{x}})/n_\gamma$   with $Q=L$ or $B$.
In equation (\ref{eq:vlal}) we have assumed  that  the  electric charge  
of the medium vanishes  ($L_e = B_p$), and that all particles
are in thermal  equilibrium with the same  temperature $T$.
The  asymmetry relevant for  $\nu_\tau \leftrightarrow \nu_s$ 
oscillations  can be obtained  from equation (\ref{eq:vlal}) simply  
interchanging  the labels $\mu$ and $\tau$;
for  $L^{(e)}$  
one has also  to include  the contribution of charged  current interactions
with electrons  adding a  term $L_e$.
The constants $a_L$ and $b_\alpha$  are:
\begin {equation}
a_L = 2 \sqrt{2} G_F \,  { 2 \zeta(3) \over \pi^2 } \simeq
 8.014 \times 10^{-12} ~  {\rm MeV}^{-2},
\end{equation}
\begin {equation}
b_{\mu(\tau)}  = 2 \sqrt{2} G_F \,  \left ( {8 \over 3}  
\,{1 \over M_Z^2} \right )
\; \left ( {\pi^2 \over 15 }\, {7 \over 8} \right ) 
\simeq   6.10  \times 10^{-21}  ~{\rm MeV}^{-4}
\end{equation}
and 
\begin {equation}
b_e  = b_\mu \, \left (1 + 2 {M_Z^2 \over M_W^2} \right )
\simeq   2.17  \times 10^{-20} ~ {\rm MeV}^{-4}.
\end{equation}

Looking  at  equations
(\ref{eq:sm})   and  (\ref{eq:vv})
we can see  that     if the   charge  asymmetry of the medium is
small, the   mixing  of neutrinos  in matter
is   strongly  suppressed:
$s^2_m \propto (T^*/T)^{12}$  for     temperatures
much  larger  than a characteristic  temperature
\begin{equation}
T^*  (\delta m^2) = 
\left ( { |\delta m^2| \over  b_{\alpha} } \right )^{1 \over 6}.
\end{equation}
Numerically we have
$T^* =23.4 ~(18.9)~|\delta m^2 ({\rm eV}^2)|^{1\over 6}$~MeV, 
for    the oscillations of $\nu_\mu$'s and  $\nu_\tau$'s ($\nu_e$'s)
with  sterile neutrinos.

\subsection {The resonance  condition}
\begin{figure} [t]
\centerline{\psfig{figure=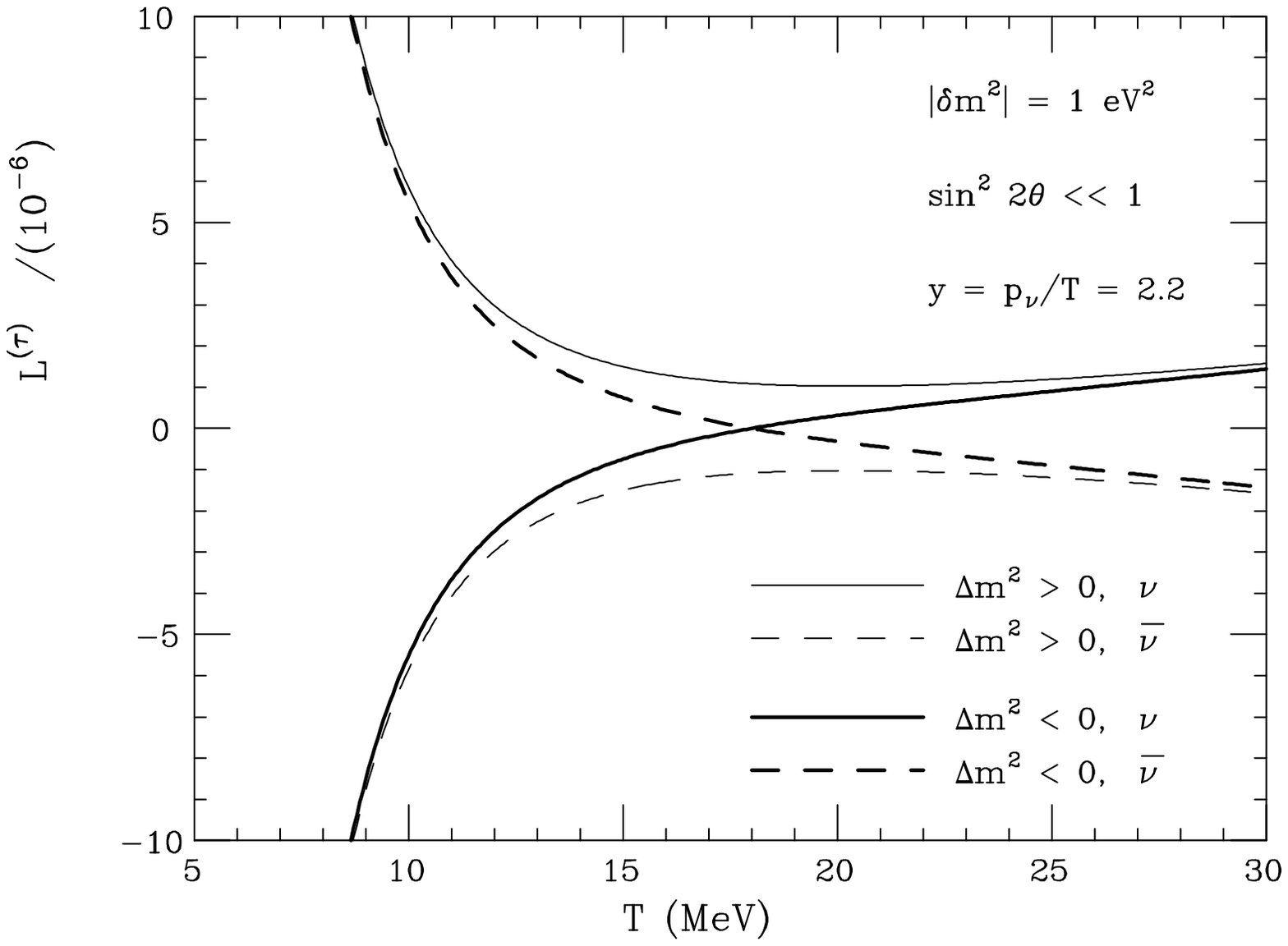,height=14cm}}
\caption {Curves in the plane ($T$,$L^{(\alpha)}$)  that 
correspond to  maximal  mixing   for neutrino  and 
anti--neutrinos  with $y = p/T = 2.2$.
All lines  are calculated for $|\delta m^2| = 1$~eV$^2$ and $s^2 \ll 1$.
\label{fig:mix}
}
\end{figure}

In the following it will be  very important to consider
which  neutrinos  (or anti--neutrinos)  have  an effective
mixing  in matter  that is  maximal.
This  is  verified   for 
$c - v_\alpha = 0$,  or  more  explicitely   for
\begin{equation}
c  \mp {a_L \, L^{(\alpha)} \,T^4 \, y \over \delta m^2}
 + {b_\alpha \,T^6 \, y^2 \over \delta m^2} = 0
\label{eq:resonance}
\end{equation}
where the upper (lower) sign applies for  neutrinos 
(anti--neutrinos).
This equation is a relation  between the  temperature $T$,
the asymmetry  $L^{(\alpha)}$ and  the   neutrino   
adimensional momentum $y = p/T$. 
As  an illustration  
in fig.~\ref{fig:mix}
we show  for  some  representative values   of the  oscillation
parameters  the    set of points 
in the plane $(T,L^{(\alpha)})$   for  which neutrinos  and antineutrinos
with  adimensional momentum $y = p/T = 2.2$ (the maximum of the
equilibrium Fermi distribution)   have  maximal  mixing.

Except  for the special case of maximal mixing in vacuum, $s^2 = 1$, 
it is  important to distinguish the cases
of  $\delta m^2 >0$   and $\delta m^2 <0$.
A  positive  (negative)  $\delta m^2$ means that 
the mass eigenstate that is  prevalently
a sterile  neutrino is  the heavier  (lighter) one.
If $\delta m^2 <0$  (and only in this case) both neutrinos
and   antineutrinos  can  have   maximal  mixing
in   an  exactly symmetric  medium  ($L^{(\alpha)} = 0$).
For  neutrinos  with   adimensional momentum $y$ this happens
at  a temperature:
\begin{equation}
\label{eq:tres}
T^0_{\rm res} (y)  =  \left ( {c \; |\delta m^2| \over b_\alpha \;y^2}
\right )^{ 1\over 6} = c^{1\over 6} \; T^*(\delta m^2) ~
y^{-{1 \over 3}}
\end{equation}
Note that  when  the  temperature  decreases,  neutrinos
with higher  and  higher  momentum $y \propto T^{-3}$ 
are resonant.
More in general (always   in the case  $\delta m^2 < 0$),
given  a  certain  value  of the  temperature $T$,
for a non vanishing    value
of the  asymmetry $L^{(\alpha)}$, 
neutrinos and anti--neutrinos have maximal mixing for
different adimensional  momenta. One has:
\begin {equation}
\label{eq:yres_}
y^{\rm res}_{\nu_\alpha (\overline{\nu}_\alpha) } = 
\pm {1 \over 2}  {a_L \over b_\alpha} \, 
{L^{(\alpha)} \over T^2}
+ {1 \over 2} \, \sqrt {
\left ( {a_L \over b_\alpha}  \, {L^{(\alpha)} \over T^2} \right )^2
+ 4 c \, { |\delta m^2| \over b_\alpha}
\; {1 \over T^6 }  }
\end{equation}
For   a  neutral medium the two  values  coincide,
for  $L^{(\alpha)} > 0$  one has 
$y^{\rm res}_{\nu_\alpha} > y^{\rm res}_{\overline{\nu}_\alpha}$, 
and  the opposite for 
$L^{(\alpha)} < 0$.
For large   asymmetries  of the  medium
the expressions  for the resonant  values of the 
adimensional   momenta  take the asymptotic  forms:
\begin{equation}
 y_{\rm high}^{\rm res} = {a_L \over b_\alpha} \;
{|L^{(\alpha)|} \over T^2}
\end{equation}
and
\begin{equation}
 y_{\rm low}^{\rm res} =  {c |\delta m^2| 
\over a_L \; |L^{(\alpha)} |\; T^4}.
\end{equation}
For  $L^{(\alpha)}$ positive 
$y_{\rm high}^{\rm res}$ applies to  neutrinos
and  
$y_{\rm low}^{\rm res}$  to  anti--neutrinos
and  vice versa for 
$L^{(\alpha)}$  negative.

In the case of $\delta m^2 > 0$,
the  situation of   maximal mixing can be  present only
for  neutrinos   or  antineutrinos  (but  not  for  both), and this is possible
only for  a  sufficiently large   charge  asymmetry of the  medium.
The  resonance    condition  is verified for  neutrinos
(antineutrinos)  only for 
$L^{(\alpha)} > L^*$
($L^{(\alpha)} < -L^*$)
where 
\begin{equation}
L^* = {2 \, \sqrt{c \;b_{\alpha}\;\delta m^2} \over a_L \,T}
\end{equation}
If this     condition is satisfied
the  neutrinos  (anti--neutrinos)  have resonant  mixing
for  two values  of the momentum given by
\begin {equation}
y^{\rm res}_{\nu_\alpha (\overline{\nu}_\alpha)} = 
+ {1 \over 2}  {a_L \over b_\alpha} \, 
{|L^{(\alpha)}| \over T^2}
\pm {1 \over 2} \, \sqrt {
\left ( {a_L \over b_\alpha}  \, {L^{(\alpha)} \over T^2} \right )^2
- 4 c \, { \delta m^2 \over b_\alpha}
\; {1 \over T^6 }  }
\end{equation}
(the  $\pm$  gives the two solutions  for  either  neutrinos  or
anti--neutrinos).

In the case of   $s^2 = 1$, 
the sign of  $\delta m^2$  is  not well defined,  and the resonant
condition  can be  obtained  as  the limit $c \to 0$  
from   both the solutions  found for 
positive  or negative $\delta m^2$. 
Effective maximal  mixing in matter is obviously present for 
both neutrinos  and  antineutrinos 
for $y = 0$,  for any   conditions   of the medium,
since in this  situation  all matter  effects  vanish;
a second  resonant  value of the
adimensional momentum 
$y^{\rm res} = (a_L/b_\mu) (|L^{(\alpha)}|/T^2)$   exists for
neutrinos  (for $L^{(\alpha)} > 0$) or
anti--neutrinos   (for $L^{(\alpha)} < 0$).

Some  examples of the   dependence on the momentum $y$
of the mixing  parameter in matter  are  shown in fig.~\ref{fig:film}, 
to be discussed in section \ref{sec:3nu}.
It is  important  to note that 
for  a   given   condition of the    medium
(fixing  the temperature  and quantum  numbers of
the plasma) the   width of the resonance, that 
is  the momentum  range 
for  which neutrinos  or  antineutrinos  
have  a large mixing  is   in general   very narrow
($\delta y/y \ll 1$) and  only a small fraction  of  all neutrino momenta
can oscillate  efficiently.
Note   also that    the  neutrinos    have  a non--negligible  
population only for values of $y$ not too far from $y_{\rm peak} \simeq 2.2$,
and therefore if the resonant  momentum is    
$y^{\rm res}_{\nu_{\alpha(\overline{\nu}_\alpha})}  \ll 0.1$
or 
$y^{\rm res}_{\nu_{\alpha(\overline{\nu}_\alpha})}  \gg 10$  the 
resonance  has   little practical  importance.

Because of the   narrowness of the   resonant  regions
it is  essential to  study the flavor    evolution of
the  neutrino ensembles  considering    in  detail 
the  momentum   distribution   of the particles.
The  approximation of  considering monochromatic  
neutrino populations   with a  single  momentum
$p = \langle p\rangle$   does  not give
a  sufficiently accurate  description  of the  problem.

\section{Kinetic equations for the sterile and active neutrino distributions}
\label{sec:QKE}

A rigorous description of the   flavor  evolution
of   the ensemble  of neutrinos  in the
early universe has to take into 
account at the same  time the  quantum coeherent  effects of  oscillations
and the   effects of  interactions with the particles of the dense  medium.
This  can be done  rigorously  making  use of a density matrix 
formalism \cite{Dolgov81}. 
The  interactions    result  in  
a loss of  quantum coherence  for  the neutrinos 
\cite{Harris82,Stodolsky87}. 
It is   possible  to write the reduced density matrix for  a  system of
two   mixed neutrino flavors  in the form
\begin{equation}
\rho(p)= 
\frac{1}{2}
\left[ P_{0}(p)I + {\mathbf P}(p)\cdot{\mathbf \sigma}\right]
\end{equation}
where $\sigma_j$ ($j=1,2,3$)  are the Pauli  matrices  and
${\mathbf P}= (P_1,P_2,P_3)$ is  a  vector.
The diagonal entries
of $\rho$ are the ratios of the classical distribution function 
($f_{\nu_{\alpha}}(p)$ for $\nu_{\alpha}$ and $f_{\nu_{s}}(p)$ for
$\nu_{s}$) with the zero chemical potential Fermi-Dirac 
distribution $f_{eq}^{0}(p)\equiv 1/[1+\exp(p/T)]$, so that:
\begin{equation}
P_{z}(p)=\frac{f_{\nu_{\alpha}}(p)-f_{\nu_{s}}(p)}{f_{eq}^{0}(p)}
  \hspace{20 mm}
P_{0}(p)=\frac{f_{\nu_{\alpha}}(p)+f_{\nu_{s}}(p)}{f_{eq}^{0}(p)}
\end{equation}
Starting from first principles it is possible to derive \cite{McKellar94}
the following set of equations  for ${\mathbf P}$ and $P_{0}$,
dubbed quantum kinetic equations (QKE):
\begin{eqnarray}\label{eq:P}
\frac{d}{dt}{\mathbf P}(\tilde{p},t)& = & 
{\mathbf V}(\tilde{p},t)\wedge{\mathbf P}(\tilde{p},t)
-D(\tilde{p},t){\mathbf P}_{T}(\tilde{p},t)+
\frac{d}{dt}P_{0}(\tilde{p},t)\hat{z}
\\ \label{eq:P_0}
\frac{d}{dt}P_{0}(\tilde{p},t) & = & R(\tilde{p},t) 
\end{eqnarray}
where:
\begin{equation}
{\mathbf P}_{T}=P_{x}\hat{x}+P_{y}\hat{y}
\end{equation}
\begin{equation}
{\mathbf V}=V_{x}\hat{x}+V_{y}\hat{y}+V_{z}\hat{z}
\end{equation}
\begin{equation}
V_{x}=\frac{\delta m^{2}}{2p}s, \hspace{5 mm}
V_{y}=0,\hspace{5 mm}
V_{z}= - \frac{\delta m^{2}}{2p}[c-v_{\alpha}(p,t)]
\end{equation}
Analogous relations hold for antineutrinos. 

The use of {\em local momenta} $\tilde{p}\equiv p \; (R/R_{*})$
($R$ is the scale factor, $R_*$ its reference value at a temperature $T_*$) 
in  the  evolution equations (\ref{eq:P})
allows to take into account the expansion of the 
universe  avoiding the explicit presence of a partial derivative term in the 
left hand sides (see for example \cite{Bernstein88}). 
As we are dealing with constant photon number, $R/R_{*}=T_{*}/T$ 
and thus we can use the convenient adimensional momentum $y \equiv p/T$.

The first term in the right hand side of equation (\ref{eq:P}) describes 
the coherent 
evolution of mixed neutrinos, while the second term describes the
decoherence effect   of interactions
through a {\em damping function} $D(y)$ 
that has the simple expression \cite{Bell99}:
\begin{equation}\label{eq:D}
D(y)=\frac{1}{2}\Gamma_{\alpha}(y) \; ,
\end{equation} 
where
\begin{equation}
\Gamma_{\alpha}(y)=
G^{2}_{F} \, k_{\alpha} ~T^{5} \, y
\label{eq:Gamma}
\end{equation} 
is the total collision rate for an active neutrino $\nu_{\alpha}$ 
with momentum $p = y\,T$
and $k_{\alpha}\simeq 0.92$   (1.27)
for $\alpha=\mu,\tau$ ($\alpha = e$).
In equation (\ref{eq:Gamma}) we  have  neglected
Pauli blocking  factors, and assumed that the  charge asymmetry of the medium
is  negligible.
Correction terms for  the presence of   asymmetries  in the 
neutrino  lepton number   are  given in \cite{Bell99}
and are small.  With this assumption the 
damping  functions  for neutrinos
and  anti--neutrinos  are equal.
It is also convenient to introduce an 
adimensional damping function:
\begin{equation}\label{eq:addf}
d_{\alpha}(y) \equiv {2\;p \over |\delta m^2|}\; D(y) =
{G_F^2 \,k_\alpha\,T^6 \,y^2 \over  |\delta m^2|}.
\end{equation} 
The third term $(dP_{0}/dt)\hat{z}$ in equation 
(\ref{eq:P})
takes into account the variation of the
total number of standard plus sterile neutrinos. This happens because, while 
$\nu_{\alpha}$ oscillate producing $\nu_{s}$, the inelastic collisions refill 
the quantum $\nu_{\alpha}$ states to preserve chemical equilibrium and this
effect changes only the $z$-component. The {\em repopulation function} $R(y)$, 
driving the evolution of $P_{0}$, for $T\gtrsim 1$~MeV  has the form 
\cite{Bell99}: 
\begin{equation}
R(y) \simeq \Gamma_\alpha (y)
\left\{\frac{f_{eq}^{\xi_{\alpha}}(y)-f_{\nu_{\alpha}}(y)} 
{f_{eq}^{0}(y)}\right\} \; .
\label{repop}
\end{equation}
In equation (\ref{repop}), 
$f_{eq}^{\xi_{\alpha}}\equiv 1/[1+\exp(y-\xi_{\alpha})]$ is
the Fermi-Dirac distribution and $\xi_{\alpha}\equiv \mu_{\alpha}/T$ is the 
adimensional chemical potential; the approximation symbol
means that the expression is rigorously true only in the limit
of complete thermal equilibrium for the non oscillating particles.
When the average collision rate is much larger than
the  expansion rate of  the universe $H(T)$, 
the deviations from  equilibrium
are so small that the term  $(dP_{0}/dt)\hat{z}$ can be neglected 
and equation (\ref{eq:P}) becomes a {\em damped Bloch equation}, well-known  
in studies of nuclear and electronic spin motion.
An important thing to notice is that the initial 
condition of the system is necessarily ${\mathbf P}_{T}=0, P_{0}=1$, because
at high enough temperatures the damping parameter would be always so
high to destroy any transversal component ({\em Turing or Zeno Paradox})
\cite{Harris82} and because the presence of sterile neutrinos is
initially negligible. In fact a sterile neutrino decouples
at very early times when $T \gg 100$~GeV and thus its density
is highly diluted by the subsequent disappearance of degrees of freedom, 
especially at the quark-hadron phase transition. 

In this  work, 
to  study the flavor evolution of the ensemble  of neutrinos, 
we  will use  a   simpler  set  of  equations that  is  given below:
\begin{equation}
\label{eq:pbo1}
\frac{d}{dt}f_{\nu_{s}}(y,t)=
\Gamma_{\alpha s}(y)\cdot
\left[f_{\nu_{\alpha}}(y,t)-f_{\nu_{s}}(y,t)\right]
\end{equation}
\begin{equation}
\label{eq:pbo2}
\frac{d}{dt}f_{\bar{\nu}_{s}}(y,t)=
\bar{\Gamma}_{\alpha s}(y)\cdot
\left[f_{\bar{\nu}_{\alpha}}(y,t)-f_{\bar{\nu}_{s}}(y,t)\right]
\end{equation}
\begin{equation}
\label{eq:pbo3}
\frac{d}{dt}f_{\nu_{\alpha}}(y,t)=
\Gamma_{\alpha s}(y)\cdot
\left[f_{\nu_{s}}(y,t)-f_{\nu_{\alpha}}(y,t)\right]
+\sum_{r}C_{r}[f_{\nu_{\alpha}}]
\end{equation}
\begin{equation}
\label{eq:pbo4}
\frac{d}{dt}f_{\bar{\nu}_{\alpha}}(y,t)=
\bar{\Gamma}_{\alpha s}(y)\cdot
\left[f_{\bar{\nu}_{s}}(y,t)-f_{\bar{\nu}_{\alpha}}(y,t)\right]
+\sum_{\bar{r}}C_{\bar{r}}[f_{\bar{\nu}_{\alpha}}]
\end{equation}
where \cite{Foot97}:
\begin{equation}\label{eq:gamma}
\Gamma_{\alpha s}(y)
=
{\Gamma_\alpha \over 2} \, s_m^2 \,  \left [1 + \left 
( {\Gamma_\alpha \ell_m \over 2}
\right )^2 \right ]^{-1} =
\frac{1}{4}
\frac{\Gamma_{\alpha}(y)\cdot s^{2}}
{s^{2}+d^{2}_{\alpha}(y)+[c-v_{\alpha}^{T}(y)-v_{\alpha}^{L}(y)]^{2}}
\end{equation}
\begin{equation}\label{eq:bgamma}
\bar{\Gamma}_{\alpha s}(y)
=
{\Gamma_\alpha \over 2} \, \overline {s}_m^2 \,  \left [1 + \left 
({\Gamma_\alpha \overline{\ell}_m \over 2}
\right )^2 \right ]^{-1} 
=\frac{1}{4}
\frac{\Gamma_{\alpha}(y)\cdot s^{2}}
     {s^{2}+d^{2}_{\alpha}(y)+[c-v_{\alpha}^{T}(y)+v_{\alpha}^{L}(y)]^{2}}
\end{equation}
(for the explicit  form   in the right--hand side  we have used
the  fact that   $s_m^2/\ell_m^2 = s^2/\ell_0^2$ and  the   definition
(\ref{eq:addf})   for $d_\alpha$).
The $C_{r}[\mbox{ }]$ in equations (\ref{eq:pbo3}) and (\ref{eq:pbo4}) 
are the collisional operators \cite{Bernstein88} for
the different reactions involving   the  standard neutrinos
and  other particles  in the medium.
This   set  of  equations  
can be rigorously derived from the quantum
kinetic equations (QKE) under an appropriate set of approximations, 
as first discussed in \cite{Shi96,Foot97}
and shown in great detail in \cite{Bell99}. 

The  equations  have 
a  simple  probabilistic  interpretation, and  
can  be derived with  heuristic   considerations \cite{Foot96,Foot97}.
Active and sterile neutrinos  
are described as two different particle species
and oscillations work as a special inelastic process that can transform an 
active neutrino into  a sterile neutrino and vice versa. 
The rate  for  the  flavor  transitions ($\alpha \to  s$ 
or $s \to \alpha$)
can be derived considering oscillations   that  develop only 
between interactions
(or  ``measurement processes'')  that  happen  with   an average
time  interval  $\tau = 2/\Gamma_\alpha$  (see \cite{Foot97}  for a more
detailed discussion).
From these considerations one  obtains:
\begin{equation}
\Gamma_{\alpha s}=\Gamma_{s \alpha}=
{\Gamma_{\alpha}(y) \over 2}  ~
\left\langle P_{m}(\nu_{\alpha}\to\nu_{s},t_{in}+\tau)\right\rangle _{\tau}
= 
{\Gamma_{\alpha}(y) \over 2}  ~
s_m^2  ~ \left [1 + \left ( {\Gamma_\alpha \, \ell_m \over 2} 
\right )^2 \right ]^{-1} \; .
\end{equation}

The solution of the   flavor  evolution equations
is  greatly simplified    assuming that the  standard  neutrinos  remain
in chemical  equilibrium.
The effect of the collisional operators for sufficiently large temperature,
i.e. for 
$T \gtrsim  3.5$ (2.5)~MeV for  $\alpha =\mu,\tau$  ($\alpha = e$),
is in fact to maintain, to a very good  approximation,  in  
thermal and chemical equilibrium
the population of standard  neutrinos.  
The  elastic  reactions 
redistribute the lepton number in a thermal way and
the inelastic processes refill continously the quantum 
states depleted by the oscillations.

The assumption   of thermal  equilibrium for the standard neutrinos 
does  not determine  completely  their    momentum  distributions, because 
 in general  $\nu$'s  and $\overline{\nu}$'s 
have  different  oscillation   probabilities; this
can generate an asymmetry   in the  lepton number $L_{\nu_\alpha}$
(that cannot  be  cancelled  by collisions) and a non  vanishing  
chemical potential.
The following relation connects the lepton number $L_{\nu_{\alpha}}$ to the 
adimensional chemical potentials for neutrinos  and anti--neutrinos:
\begin{equation}
L_{\nu_{\alpha}}=\frac{1}{n_{\gamma}}
\int\frac{d^{3}p}{(2\pi)^{3}}
\left(\frac{1}{e^{y-\xi_{\alpha}}+1}-\frac{1}{e^{y-\bar{\xi}_{\alpha}}+1}\right)
\; ,
\label{eq:xi}
\end{equation}
which for $\xi_{\alpha},\bar{\xi}_{\alpha}\ll 1$ becomes simply:
\begin{equation}\label{eq:Lxi}
L_{\nu_{\alpha}}=\frac{\pi^{2}}
            {24\mbox{ }\zeta(3)}(\xi_{\alpha}-\bar{\xi}_{\alpha})
+{\cal O}(\xi_{\alpha}^{2},\overline{\xi}_\alpha^{2})
\end{equation}

If the standard  neutrinos  are  in thermal  equilibrium 
one  has
$\bar{\xi}_{\alpha}=-\xi_{\alpha}$ and the active neutrino distributions are 
described only through $L_{\nu_{\alpha}}$. 
Below the chemical decoupling 
temperature this is not true in general and 
one should take into account also the quantity 
$\xi_{\alpha}$ + $\bar{\xi_{\alpha}}$:
this is important if one wants to describe the effects of electron
chemical potentials on BBN \cite{Foot97b}. 
For the temperatures relevant for our  discussion chemical equilibrium
is verified, and therefore  
these  effects  are   not important
and  we will make the   approximation that a
single quantity
$L_{\nu_\alpha}$  fully determines  the  momentum  distributions  of
both neutrinos an anti--neutrinos  of   flavor $\alpha$.

The  simplified framework we  have  discussed
and  that we will will use  in this  work
gives a  good description  of the flavor
evolution of  neutrinos  in the  early  universe 
only if some conditions  are  satisfied.
A detailed discussion  of these  conditions 
and  a formal  derivations  of    the Pauli--Boltzmann 
equations     from  the full  set of Quantum Kinetic Equations 
 for the case of two--flavor oscillations of one 
active and one sterile neutrino
can  be  found in \cite{Bell99}.
Here we will recall these  conditions 
including  some  simple  comments.
There is a small angle condition, that is simply
\begin{equation}
s_{m}^{2}\ll 1 \; ,
\label{eq:small1}
\end{equation}
or, when the  effective mixing in matter is  large
(that is  close to a resonance) 
the   condition
\begin{equation}
{\ell_{\rm int} \over \ell_m} = {\sqrt{s_\alpha^2 + (c_\alpha - v_\alpha)^2}
\over d_\alpha } \ll 1.
\label{eq:small2}
\end{equation}
Each one of these conditions   assures that  
all  neutrinos in the    early universe  medium  remain   very close
to  a pure   flavor  (standard or sterile) eigenstate.
This is  obvious  in the case of  small  effective mixing  in matter;
the condition  (\ref{eq:small2})  assures  that
frequent interactions interrupt 
the  coherent   flavor evolution, continuously reprojecting the neutrino
states  into  flavor  eigenstates.
Therefore   if the conditions (\ref{eq:small1}) or (\ref{eq:small2}) are
satisfied the  transverse components of
the vectors ${\mathbf P}$ and ${\mathbf {\overline P}}$ 
that describe  coherent  superposition of  flavor eigenstates 
remain always  small, the  number  of degrees of freedom is  reduced,
and the Pauli--Boltzmann equations  become  a good  description of the 
evolution of   neutrino   populations.

An additional  requirement  for the validity
of the  framework  considered 
is that   the relevant properties  of the medium 
(such as the effective  potential and  the damping  function) 
do  not   change appreciably in the time $\ell_{\rm int}$ between
collisions,
otherwise the diagonal terms of the density matrix 
have still an appreciable dependence on the derivatives of the off-diagonal 
terms. In \cite{Bell99} it is shown that the parameter variation is maximal
at the resonance and that 
the (most restrictive) condition assumes the form $D\gtrsim 6H$.
This condition implies that in order to be confident in the approximation
the bulk of (anti--)neutrinos should enter the resonance at a temperature
$T_{\rm res} \gtrsim 3\mbox{ } {\rm MeV}\;$.
If the charge number is negligible this implies 
that the Pauli-Boltzmann approach 
can be used only for $|\delta m^{2}|>10^{-5}$~eV$^2$,
see equation (\ref{eq:tres}).
For $T_{\rm res}  \lesssim 3$~MeV, full MSW conversion can occur at the 
resonance, 
an effect that the Pauli-Boltzmann approach neglects. In this situation major
differences from a full quantum approach are expected.

We note that even in presence of an appreciable transversal 
component   of the vectors ${\mathbf P}$ and ${\mathbf {\overline P}}$ 
approximately correct results can be obtained (provided that full 
MSW conversion does not occur) 
for the calculation of sterile neutrino 
contribution to the effective 
number of neutrinos plus antineutrinos. 
In fact the 
motion of the vectors around the ${\mathbf V}$ vector 
affects much more the difference (lepton asymmetry) than the sum.
For the lepton asymmetry evolution the differences are 
more relevant and a classic approach misses the oscillatory behaviour at 
large angles \cite{Enqvist91,Shi96,Foot99} 
(of course below 3~MeV it also neglects 
MSW conversion effect on the lepton asymmetry evolution).
Aiming at the evaluation of the effective neutrino plus antineutrino 
number it is meaningful to extend the classic approach also
to high values of the mixing angle. To this purpose we will keep the
term $s^{2}$ in the denominators
of equations (\ref{eq:gamma}) and (\ref{eq:bgamma}). 
In \cite{Bell99}, proceeding directly 
from QKE, expressions for the rates that do not
contain this term have been obtained. As we will see
in next section, this would spoil the nice
agreement of the results that we obtain using a Pauli-Boltzmann approach
with those obtained using the QKE.

\section{Sterile $\nu$ production   for constant $L^{(\alpha)}$. }
\label{sec:prod}

In the  previous  section we have  reduced
the problem  of the   evolution of  the 
ensemble   of  neutrinos   in  the early   universe to the
solution of a  set of coupled differential equations.
The     flavor  transition rates  $\Gamma_{\alpha s}$ 
and  $\overline{\Gamma}_{\alpha s}$ 
depend  on the    lepton  number  $L_{\nu_\alpha}$
that is    determined  by  the oscillation  themselves,
and therefore we have  a non--linear problem.
In this    section we   will begin  solving a
simpler  problem 
considering the asymmetry 
$L_{\nu_\alpha}$ as  constant.
We will show  that in the   framework of our method,
the solution  of this problem  can  be  expressed in
a  very  simple  form.
In the next section we will  
discuss   the  complete  problem  and  consider the lepton
number  as a dynamical  variable.

The first step for the calculation is to rearrange the equations in a more 
convenient form, replacing the distribution functions with the 
following quantities: 
\begin{equation}
z_{s}\equiv \frac{f_{\nu_{s}}}{f_{eq}^{0}}
  \hspace{20 mm}
\bar{z}_{s}\equiv \frac{f_{\bar{\nu}_{s}}}{f_{eq}^{0}} 
\end{equation} 
As we will deal with small chemical potentials 
($\overline{\xi}_\alpha,\xi_{\alpha} \ll 1$), we can 
perform the expansion:
\begin{equation}\label{eq:xiexp}
f_{\nu_{\alpha}}(y)=
f_{eq}^{0}\left(1+\xi_{\alpha}\frac{e^{y}}{e^{y}+1}+{\cal O}(\xi_{\alpha}^{2})\right)
\end{equation}
Moreover, considering that the equilibrium distribution $f_{eq}^{0}$
satisfies the Boltzmann equation
(see e.g. \cite{Bernstein88}):
\begin{equation}
L[f_{eq}^{0}] \equiv 
{\partial \over \partial t} f_{eq}^{0}(p,t)
-pH \,
{\partial \over \partial p} f_{eq}^{0}(p,t)\equiv
\frac{d}{dt}f_{eq}^{0}(y,t)=0,
\end{equation}
it is easy to derive from equations (\ref{eq:pbo1}) and (\ref{eq:pbo2}), 
the following equations for $z_{s}(y,t)$, $\bar{z}_{s}(y,t)$
(neglecting terms ${\cal O}(\xi_{\alpha}^{2})$):
\begin{equation}\label{eq:z}
\frac{d}{dt}z_{s}(y,t)=
\Gamma_{\alpha s}(y,t)\left[1-z_{s}(y,t)+\xi_{\alpha}(t)\frac{e^{y}}{e^{y}+1}\right]
\end{equation} 
\begin{equation}\label{eq:bz}
\frac{d}{dt}\bar{z}_{s}(y,t)=
\bar{\Gamma}_{\alpha s}(y,t)
\left[1-\bar{z}_{s}(y,t)+\overline{\xi}_\alpha(t)\frac{e^{y}}{e^{y}+1}\right]
\end{equation}

We are interested to calculate the contribution of sterile neutrinos to the 
energy density of the universe.  For this  purpose it is  usual  to define
a sterile neutrino effective number:
\begin{equation}
N^{\rm eff}_{\nu_{s}} \equiv 
\frac{\rho_{\nu_{s}}+\rho_{\bar{\nu}_{s}}}{2\, \rho_{eq}^{0}}
\end{equation} 
where $\rho_{eq}^{0}=(7/8)(\pi^{2}/30)T^{4}$  is the energy density
per  degree of  freedom of  a  fermion  particle
in thermal  equilibrium with vanishing   chemical  potential.
The  results on  BBN   can  be used to put  constraints on this  
quantity  at a  temperaure    $T \lesssim 1$~MeV. 
The sterile neutrino effective number can be expressed as:
\begin{equation}
N^{\rm eff}_{\nu_{s}}=
\frac{120}{7\pi^{4}}\int dy \frac{y^{3}}{1+e^{y}} ~z^{+}_{s}(y)
\end{equation} 
where  $z^{+}_{s}\equiv (z_{s}+\bar{z}_{s})/2$.
It is  also  interesting to consider the
sterile neutrino charge number
$L_{\nu_{s}}\equiv (n_{\nu_{s}}-n_{\overline{\nu}_{s}})/n_{\gamma}$:
\begin{equation}
L_{\nu_{s}}=
\frac{1}{2\zeta(3)}\int dy \frac{y^{2}}{1+e^{y}} ~z^{-}_{s}(y)
\end{equation}
where  $z^{-}_{s}\equiv (z_{s}-\bar{z}_{s})/2$.
Thus we see that the interesting quantities to be calculated are 
$z^{+}_{s}$ and $z^{-}_{s}$, also because, when dealing with the coupled 
problem, these will be the quantities entering the equation for the 
$\alpha$-neutrino lepton number.
 We can derive from equations (\ref{eq:z}) and (\ref{eq:bz}) the evolution 
equations for $z^{+}_{s}$ and $z^{-}_{s}$, 
using $\xi_{\alpha}=-\overline{\xi}_\alpha$:
\begin{equation}
\frac{dz^{+}_{s}}{dt}=\frac{\Gamma_{\alpha s}+
\bar{\Gamma}_{\alpha s}}{2}[1-z^{+}_{s}]+
\frac{\Gamma_{\alpha s}-\bar{\Gamma}_{\alpha s}}{2}
\left[\xi_{\alpha}\frac{e^{y}}{e^{y}+1}-z^{-}_{s}\right]
\label{eq:zp}
\end{equation} 
\begin{equation}
\frac{dz^{-}_{s}}{dt}=
\frac{\Gamma_{\alpha s}-\bar{\Gamma}_{\alpha s}}{2}[1-z^{+}_{s}]+
\frac{\Gamma_{\alpha s}+\bar{\Gamma}_{\alpha s}}{2}
\left[\xi_{\alpha}\frac{e^{y}}{e^{y}+1}-z^{-}_{s}\right]
\label{eq:zm}
\end{equation}
From equations (\ref{eq:gamma}) and (\ref{eq:bgamma}) 
and using  the definition $v^L_\alpha = \tilde {v}(y,T) \,L^{(\alpha)}$,
is easy to find the following 
expressions:
\begin{equation}\label{eq:gmbg}
\Gamma_{\alpha s}-\bar{\Gamma}_{\alpha s}=
\frac{s^{2}\Gamma_{\alpha}}{\Delta}\cdot
\tilde{v}(c-v^{T}_{\alpha})\cdot L^{(\alpha)}\; ,
\end{equation}
\begin{equation}\label{eq:gpbg}
\Gamma_{\alpha s}+\bar{\Gamma}_{\alpha s}=
\frac{1}{2}\frac{s^{2}\Gamma_{\alpha}}{\Delta}\cdot
[s^{2}+d^{2}_{\alpha}+(c-v^{T}_{\alpha})^{2}+(v^{L}_{\alpha})^{2}]\; ,
\end{equation}
with 
\begin{equation}
\Delta\equiv
[s^{2}+d^{2}_{\alpha}+(c-v^{T}_{\alpha}-v^{L}_{\alpha})^{2}]\cdot
 [s^{2}+d^{2}_{\alpha}+(c-v^{T}_{\alpha}+v^{L}_{\alpha})^{2}] \; .
\end{equation}
 These equations allow to show that the terms explicitly containing 
$\xi_{\alpha}$ can be safely neglected: 
in the equation (\ref{eq:zp}) the term containing 
$\xi_{\alpha}$  is actually an 
${\cal O}(\xi_{\alpha}^{2})$ term, 
while in the equation (\ref{eq:zm}) the analogous term gives a 
small correction because  
$[s^{2}+d^{2}_{\alpha}(y,T)+(c-v^{T}_{\alpha}(y,T))^{2}+(v^{L}_{\alpha})^{2}]\ll
\tilde{v}(y,T)[c-v^{T}_{\alpha}(y,T)]$ for the relevant values of $y$ and
in the range of temperatures we are considering.
Thus we can neglect the $\xi_{\alpha}$ terms in 
the equations for $z_{s}$ (\ref{eq:z})
and $\bar{z}_{s}$ (\ref{eq:bz}).

It is  convenient to adopt as  independent variable in these  
equations the temperature instead of the time,
using:
\begin{equation}
{dT \over dt} = - H(T) \, T = - \; {\alpha_H \, T^2 \over M_P } \; T
\label{eq:relTt}
\end{equation}
where $M_P$ is the  Planck mass and
\begin{equation}
\alpha_H = \sqrt{ {\pi^2 \over 30} \, {43\over 4} \, {8 \pi \over 3} }
\simeq  5.44 \; .
\end{equation}
One obtains in this way \cite{Foot97}:
\begin{equation}
\frac{d}{dT}z_{s}(y,T)=
-\frac{\Gamma_{\alpha s}(y,T)}{H(T)T}\left[1-z_{s}(y,T)\right] 
\end{equation}
\begin{equation}
\frac{d}{dT}\overline{z}_{s}(y,T)=-\frac{\bar{\Gamma}_{\alpha s}(y,T)}{H(T)T}
                       \left[1-\bar{z}_{s}(y,T)\right] \; .
\end{equation}
 These equations are now simple enough to allow analytic solutions  
for the different parameters involved (lepton number included).
It is  convenient  to write the  solution in the  following form
\cite{Cline92}:
\begin {equation}
z_{s} (y,T) \equiv {f_{\nu_s} (y, T) 
\over f^0_{eq} (y) } = 1
- \exp \left [ - g_s (y, T) \right ].
\end{equation}
The  function   $g_s (y,T)$,
assuming  that  the asymmetry is 
constant  in time, 
can be calculated  with a simple  integral:
\begin {equation}
g_s (y, T)   = 
\int_{T}^{\infty}dT'~\frac{\Gamma_{\alpha s}(y,T')}{H(T')T'}.
\end{equation}
For $\overline{z}_s$ and 
 ${\overline g}_s$ one  has simply  to  make the substitution
$\Gamma_{\alpha s}  \to \overline{\Gamma}_{\alpha s}$.
Using the definitions of
$\Gamma_{\alpha s}$ and $H(T)$  one can 
write  more explicitly:
\begin {equation}
g_s (y, T) = 
K_\alpha \, |\delta m^2|^{1 \over 2} \, s^2 ~
G_{\nu_s} (y, T) \; ,
\label{eq:g0}
\end{equation}
with  a similar  expression   for  anti--neutrinos.
In equation (\ref{eq:g0}) we defined 
\begin{equation}
K_\alpha =  {G_F^2 k_\alpha \,M_P \over 4\,b_\alpha^{1\over 2}\, \alpha_H}
\simeq  898\;(657)~{\rm eV}^{-1}
\end{equation}
for $\alpha = \mu,\tau$ $(\alpha = e)$, 
and   the function $G_{\nu_s} (y,T)$, 
that  depends    on the  parameters  $s^2$, $\delta m^2$ and
$L^{(\alpha)}$   :
\begin {equation}
G_{\nu_s} (y, T) =
\int_{T/\overline {T} (y, \delta m^2)}^\infty \, dt ~
{t^2  \over s^2 +  d_0^2\,t^{12} + 
\left [ c \mp   a_L \; b_\alpha^{-{2\over 3}}\; L^{(\alpha)} 
|\delta m^2|^{-{1\over 3}} \, y^{-{1\over 3}} \, t^4
\pm t^6 \right ]^2 }
\label{eq:g-all}
\end{equation}
In equation (\ref{eq:g-all}) the upper (lower) signs  hold for $\delta m^2 >0$ 
($\delta m^2 < 0$),
and  the   constant $d_0$ is given by 
\begin {equation}
d_0 =  {G_F^2 \,k_\alpha \over b_\alpha} \simeq 0.02\,(0.008)\ \ {\rm for}\; 
\alpha=\mu,\tau\;(\alpha = e)\; .
\end{equation}
The  function  $G_{\overline{\nu}_s}$ is obtained 
changing the sign of the term   containing $L^{(\alpha)}$.
In  the  lower  limit of integration 
we have introduced  the definition:
\begin{equation}
\overline {T} (y, \delta m^2) = 
\left ( { |\delta m^2| \over b_\alpha \, y^2 } \right )^{ {1\over 6} }\; .
\end{equation}

We are interested in the limit of small $T$, i.e. generally 
$T < \overline {T} (y, \delta m^2)$ (for typical $y$ values).
Note that for   $\delta  m^2 < 0$ and in the absence of
asymmetry, neutrinos  and anti--neutrinos
with adimensional momentum $y$ have  maximal  mixing
at $T = c^{1\over 6}\, \overline {T}  (y, \delta m^2)$,
see equation (\ref{eq:tres}).

\subsection {Vanishing   charge asymmetry  of   the medium}
\label{ss:4.1}
It is  interesting  to consider  first the case where
the   charge asymmetry   of the medium is   negligibly  small.
In this case the  population of  sterile  neutrinos
and  anti--neutrinos   created  by the oscillations
are  identical.  
For $L^{(\alpha)} = 0$
equation (\ref{eq:g-all}) simplifies to 
\begin {equation}
G_{\nu_s (\overline{\nu}_s)}^\pm (y, T) =
\int_{T/\overline {T} (y, \delta m^2)}^\infty \, dt ~
{t^2  \over s^2 +  d_0^2\,t^{12} + 
( c  \pm t^6  )^2 }
\label{eq:g-l0}
\end{equation}
where  the positive (negative) sign  refer to 
$\delta m^2 >0$ ($\delta m^2 <0$), while neutrinos  and anti--neutrinos
have the same  distribution.
Note that all  dependence  on  $|\delta m^2|$  and  $y$ is   in the value
of  the lower limit   of integration.   For  $T\to 0$  the dependences
on $y$ and $|\delta m^2|$  disappear  and  the function $G$   depends only on 
the mixing  parameter  $s^2$.
\begin{equation}
 \lim_{T \to 0}\; G_{\nu_s (\overline{\nu}_s)}^\pm (y, T) 
= F_\pm (s^2) = \int_0^\infty
dt ~{t^2 \over s^2  + d_0^2 t^{12} + (c \pm t^6)^2 }
\end{equation}
the $\pm$ sign  again refers to the
cases  of positive or negative $\delta m^2$.

\begin{figure} [t]
\centerline{\psfig{figure=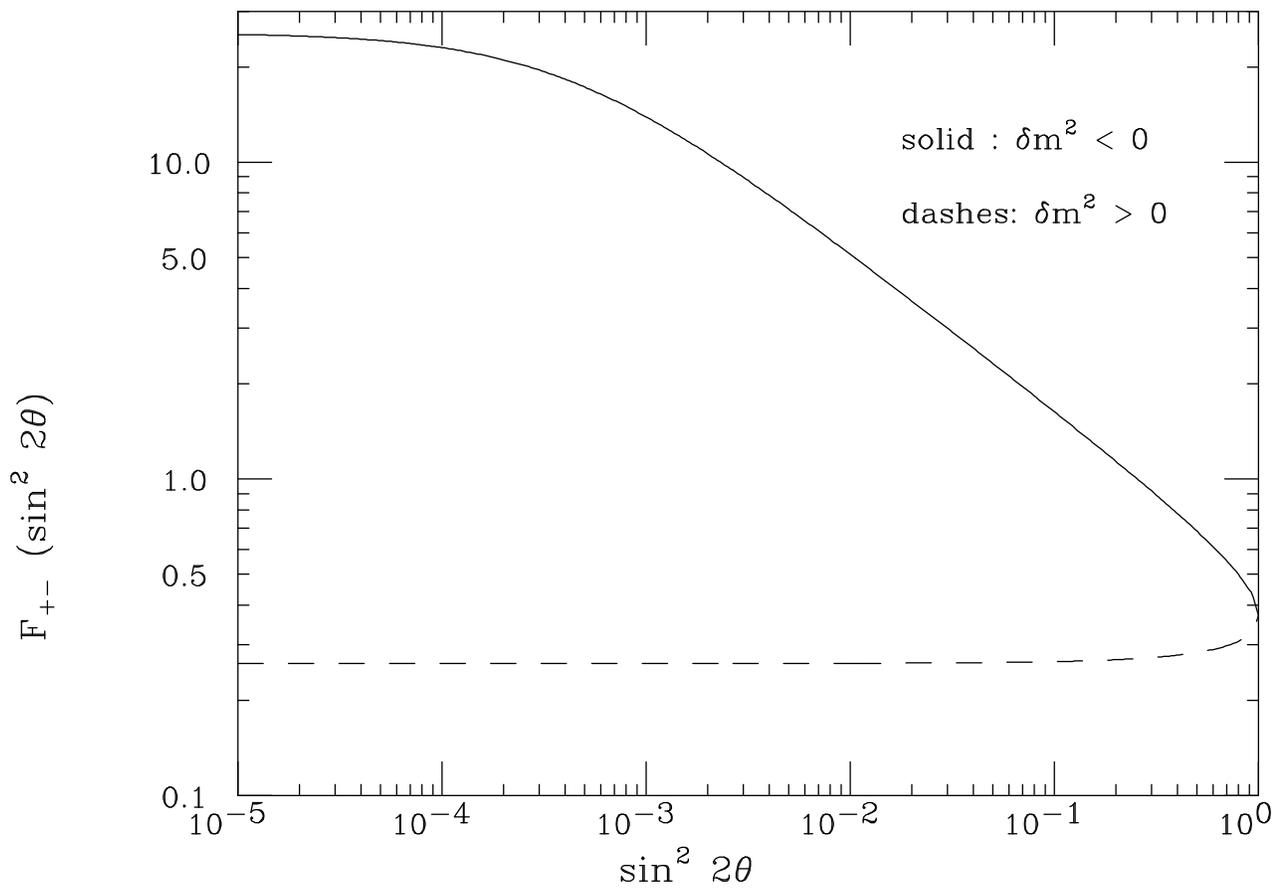,height=14cm}}
\caption {Plot of the functions $F_+(s^2)$ 
(dashed line)  and
$F_-(s^2)$  (solid  line)  as  a  function of
the mixing parameter.
\label{fig:ff}
}
\end{figure}

In fig.~\ref{fig:ff}  we plot the  functions $F_{\pm}(s^2)$. 
For maximal  mixing  the     two    curves  have 
the same value ($F_{\pm}(1) \simeq \pi/(6\,\sqrt{2}) \simeq 0.37$), 
while for   $s^2\lesssim 10^{-4}$
 the  two functions 
tend  to    values  that  differ by approximately two orders  of magnitude
($F_-(0) \simeq 25.3$, $F_+(0) \simeq 0.26$).  This  is   easily  understood  
noting that for  $\delta m^2 <0$ 
and  $L^{(\alpha)} = 0$, both  neutrinos  and  anti--neutrinos
with adimensional  momentum $y$ 
go through a resonance for a brief time interval when the 
temperature   is  close  to the value 
$c^{1\over 6}\, \overline {T}  (y, \delta m^2)$,
while in the case of positive $\delta m^2$ the resonance  is  absent.

\begin{figure} [t]
\centerline{\psfig{figure=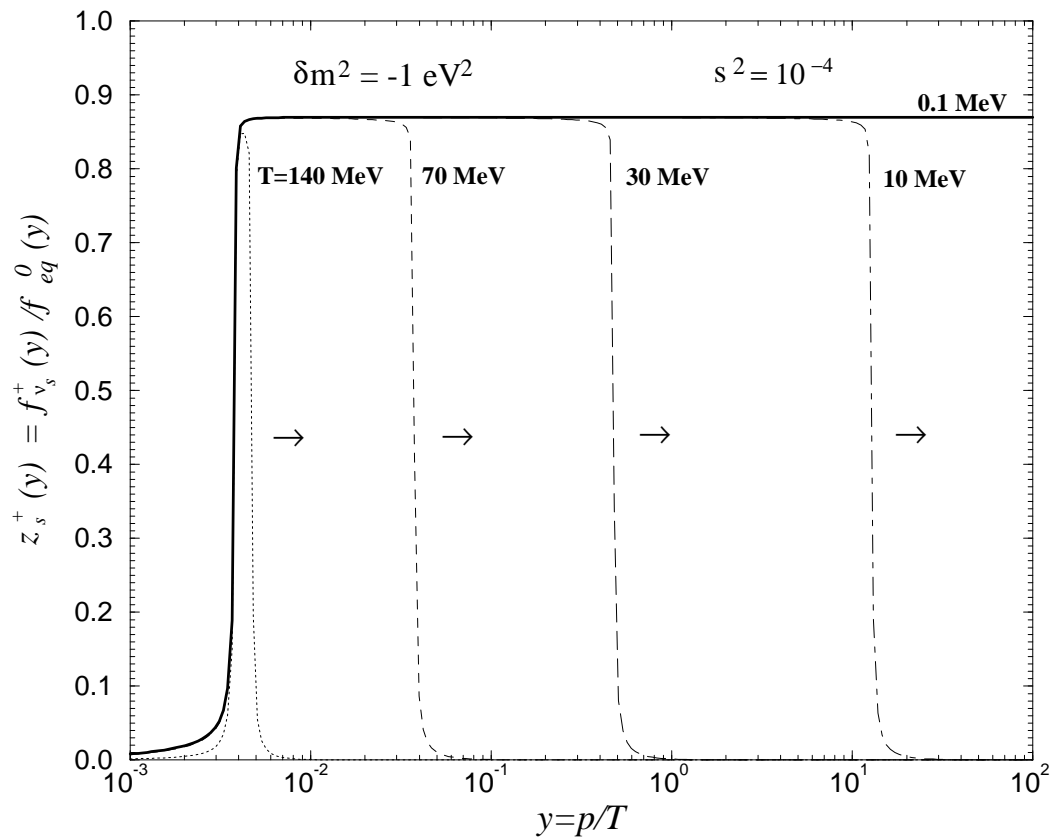,height=16cm,angle=270}}
\caption {
Evolution of the sterile neutrino distribution
assuming the presence of $\nu_s\leftrightarrow \nu_{\mu(\tau)}$  oscillations
with $\delta m^2  = -1$~eV$^2$ and 
$s^2 = 10^{-4}$.   The    flavor  evolution  is   followed
from  an initial  temperature $T_i = 150$~MeV, where  we assumed 
that  sterile neutrinos  are  absent.
We  plot the ratio of momentum distribution 
of $\nu_s$'s  and ${\overline{\nu}}_s$'s  (equal in this case) with 
a thermal equilibrium distribution with zero chemical potential.
The thick  solid  line  shows the   distribution at $T = 0.1$~MeV,
the thin dot--dashed, long--dashed  and dashed   lines
 show the distributions   for 
higher  values of $T$.
At each temperature the production of $\nu_s$ 
occurs only in a small interval of momenta
where  oscillations  are  resonant. As an illustration  
the dotted curve shows the region  where sterile neutrinos
are produced  for  $T$  just below  150~MeV.
The direction of the arrows 
indicates how the resonant momentum is changing with time. 
\label{fig:evol0}
}
\end{figure}

Most  of the production of  the  sterile  neutrinos
for the case $\delta m^2 < 0$ happens  during the short time
when  maximal  mixing  is   produced.
This is illustrated in 
fig.~\ref{fig:evol0}    that  shows one example of
the evolution  of the sterile neutrino    population 
for negative $\delta m^2$. 
The  fact that the  final    momentum  distribution 
of the sterile  neutrinos  is 
proportional to a   thermal  distribution is
the non trivial   effect of an  exact   cancellation 
between the amount of time  that  each  momentum 
spends  at the resonance, and the degree of damping of the oscillations
due to  the interactions.

Note that    in the plane
($s^2$, $\delta m^2$), a  line of constant  $N_{\nu_s}^{\rm eff}$
can be obtained  from 
\begin {equation}
N_{\nu_s}^{\rm eff} = 1 - \exp \left [ -K_\alpha  \, |\delta m^2|^{1 \over 2}
\,s^2 \,F_\pm (s^2) \right ] \; .
\end{equation}
For  small  $s^2$  the corresponding line is:
\begin{equation}
\label{eq:constr}
|\delta m^2|^{1 \over 2} \; s^2 = 
 -{ 1\over K_\alpha \;F_\pm(0) } 
~\log [1 - N_{\nu_s}^{\rm eff} ] \; .
\end{equation}

\begin{figure} [t]
\centerline{\psfig{figure=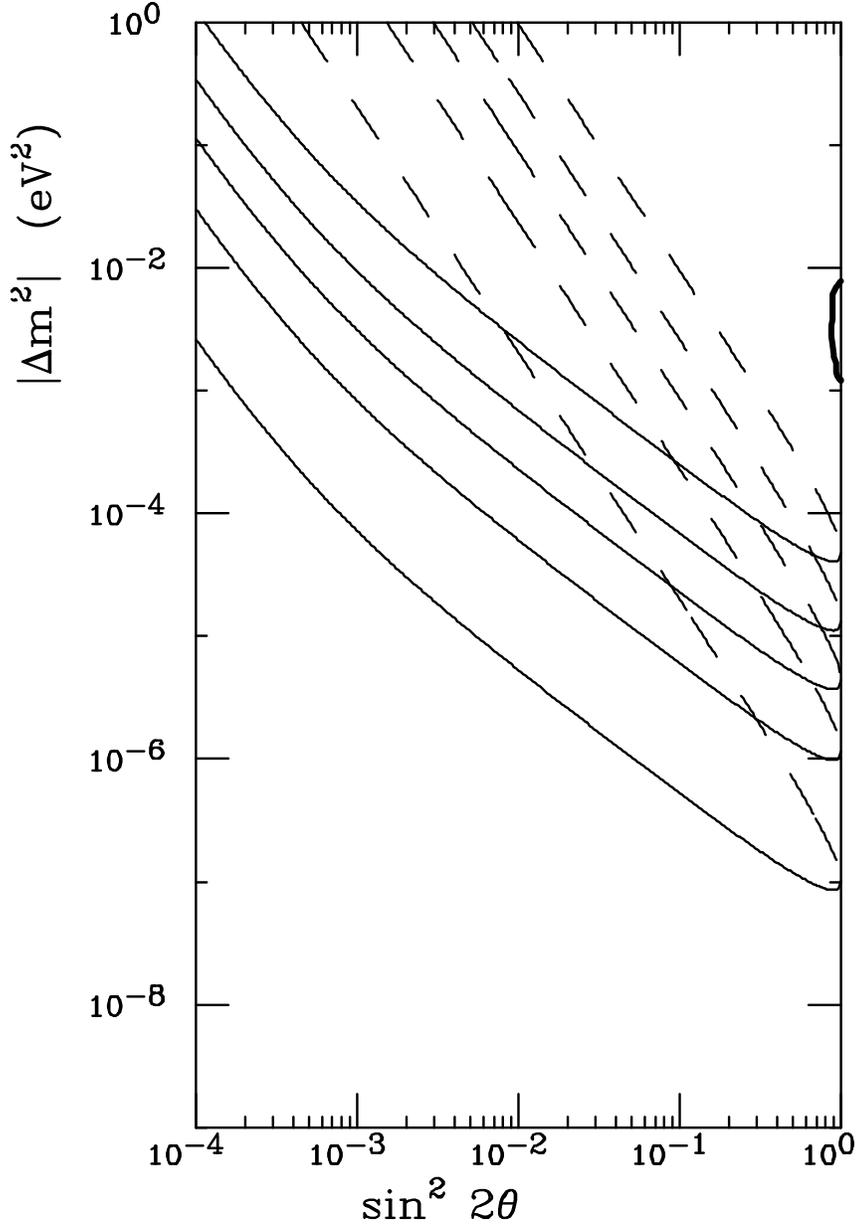,height=18cm}}
\caption {Lines in the plane  ($s^2$,$|\delta m^2|$)
that  correspond  to  a fixed   value of 
the sterile  neutrino effective  number  
$N_{\nu_s}^{\rm eff}$: 0.1, 0.3, 0.5, 0.7, 0.9 (from bottom up).
The  curves are calculated  assuming that the asymmetry of the medium  
remains negligibly small.
The solid  (dashed) lines
correspond to $\delta m^2 <0$ 
($\delta m^2 > 0$).
The  thick  solid  line  shows 
the allowed  region  indicated by
the Super--Kamiokande experimental results.
\label{fig:lim}
}
\end{figure}

In fig.~\ref{fig:lim}   we  show the  curves in the plane
$(s^2,|\delta m^2|)$    that  correspond to a constant  value 
$N_{\nu_s}^{\rm eff}$  that is 
proportional to the energy density   of the sterile  neutrinos
at low temperature.
It can  be  seen  that the region   of  parameters  that is  allowed
by the  atmospheric  neutrino  data  implies
that the sterile  neutrinos   are    fully thermalized 
(with $N_{\nu_s}^{\rm eff}$    very  close  to 1)
and this is possibly incompatible with bounds \cite{BBN} from
primordial nucleosynthesis. We note that this conclusion follows
from the assumptions that the asymmetry of the medium is negligibly
small and that the oscillations only happen between a sterile and
a single active neutrino (and their antiparticles).

Our analytic results improve those from the rough
criterium $\Gamma_{\alpha s}/H(T) \lesssim 1$ 
\cite{Fargion84,Barbieri90,Kainu90,Enqvist92b,Shi93}
and can be compared directly with 
the numerical results from density matrix approach found in
\cite{Enqvist92b,Shi93} (where momentum-dependence was not included, but
it appears from our results that it does not play any role). 
In the case of interest for us, $\alpha=\mu,\tau$, the comparison gives a
 good agreement. It may be noted that for maximal mixing we have 
slightly too restrictive results since we neglected the effect of 
depletion of active neutrino numbers when transitions occur below the 
chemical decoupling temperature at about $3.5$ $\rm Mev$.
 For this same reason our results cannot be applied to 
the case $\alpha=e$ for $\delta m^{2}<10^{-5}$. In this case the depletion 
effect must be taken into account due to the direct role of electron
neutrinos in the neutron-proton interconversion rates and moreover the MSW 
transition plays a fundamental role. 

\subsection {Non vanishing charge  asymmetry}
\label{ss:4.2}
\begin{figure} [t]
\centerline{\psfig{figure=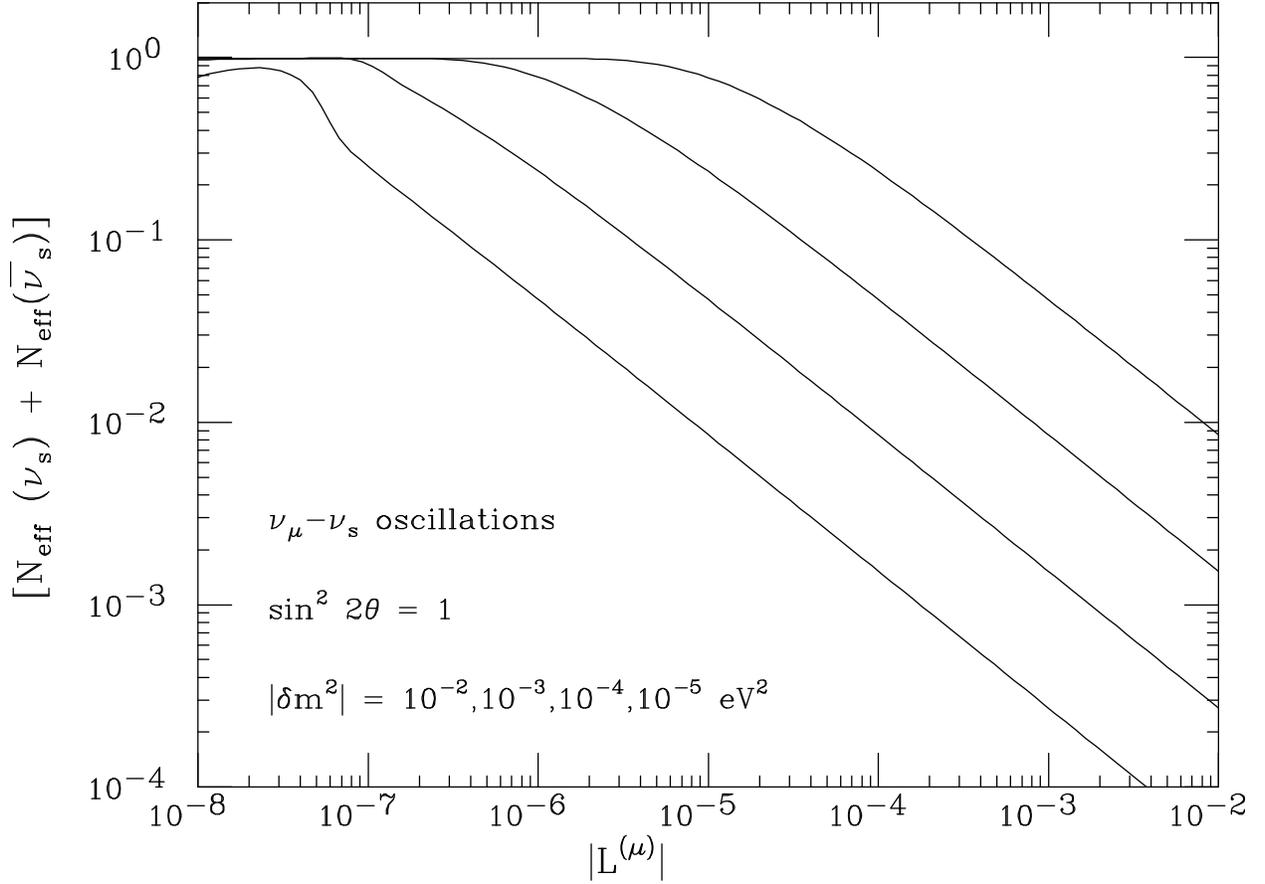,height=14cm}}
\caption {Plot of the effective number   of  sterile neutrinos
$N_{\nu_s}$(eff)    as a function of  $L^{(\mu)}$  
calculated   assuming  that the sterile  neutrino
is  maximally mixed  with a muon
(or  a tau)  neutrino.  The  curves are  calculated  for 
$|\delta m^2| = 10^{-2}$, $10^{-3}$, $10^{-4}$ and $10^{-5}$~eV$^2$
(from right to left).
In the calculation the asymmetry $L^{(\mu)}$ is   fixed
and it is not  considered  as  a  dynamical  variable.
\label{fig:asym1}
}
\end{figure}

We consider now the case where a non vanishing 
(constant) asymmetry is present in the medium.  
In this case   the  density  of 
sterile  neutrinos  and  anti--neutrinos   produced  by the oscillations
could be generally different, and in turn also the 
net  lepton number of standard  neutrinos will not stay constant: it is
reasonable to approximate it with a constant if it is varying 
sufficiently slowly, and in any case the complete, dynamical calculation 
will be considered in section \ref{sec:lae}.
In fig.~\ref{fig:asym1}  we  plot the effective  number of  sterile neutrinos 
at  low  temperature   as  a function of the  (constant)
asymmetry   calculated  assuming  maximal  mixing  and for
different  values of  $|\delta m^2|$.   It can  be seen 
that   for   sufficient   large  values of $L^{(\mu)}$  the 
contribution of the  sterile   neutrinos to the  energy  density
begins  to  drop  following a power law  $N_{\nu_s}^{\rm eff} \propto 
[L^{(\mu)}]^{-{3 \over 4}}$.
From  the  figure    we can see  that with  the oscillation  parameters 
indicated  by Super--Kamiokande 
the contribution   of sterile
neutrinos to the   energy   density would be sufficiently small 
if  $L^{(\mu)} \gtrsim 10^{-4}$.

It  is not difficult to obtain  analytically an  understanding of the 
suppression of the oscillations due to  the presence 
of the asymmetry $L^{(\mu)}$.
For  $s^2 = 1$, 
the denominator  of the integrand in equation (\ref{eq:g-all})  can 
be  rewritten:
\begin{equation}
1 + d_0^2 \,t^{12} + t^8\, [|L^{(\mu)}|/\tilde{L}\pm t^2]^2
\end{equation}
where  the  negative  sign   applies to the case 
of  neutrinos   for $L^{(\mu)} > 0$,
to antineutrinos   for $L^{(\mu)} < 0$,  
and we  have  defined
\begin{equation}
\tilde{L} = {b_\mu^{2 \over 3} \over a_L} \;
|\delta m^2|^{1\over 3} \;y^{1 \over 3} \; .
\end{equation}

The results discussed in section \ref{ss:4.1} continue to hold
if $|L^{(\mu)}| \ll \tilde{L}$; we consider here the opposite case, 
$L^{(\mu)} \gg \tilde{L}$ (or $- L^{(\mu)} \gg \tilde{L}$).
In this case the function 
$G_{\nu_s (\overline{\nu}_s)} (y, T\to 0)$ -- see equation (\ref{eq:g-all}) --
receives important contributions from two different regions:
one is the region of low temperatures, 
$t \sim (L^{(\mu)}\,/\,\tilde{L})^{-1/4}$, another is a high temperature 
region, where a resonance occurs for $t = (L^{(\mu)}\,/\,\tilde{L})^{1/2}$.
It happens that the first region (low temperature) gives the dominant
contribution (and therefore the productions of sterile neutrinos and 
antineutrinos are similar), and this explains the behavior of the sterile 
neutrino number with increasing lepton asymmetry, visible in 
fig.~\ref{fig:asym1}, namely 
\begin {equation}
\left . G_{\nu_s}^{\rm low} (y, T\to 0) \right |_{s^2 = 1} \simeq 0.43\;
(L^{(\mu)}\,/\,\tilde{L})^{-3/4} \; .
\label{eq:Glow}
\end{equation}
The resonant high temperature contribution would give instead a behavior
\begin {equation}
\left . G_{\nu_s}^{\rm high} (y, T\to 0)  \right |_{s^2 = 1} 
\simeq
{1 \over 2 \,d_0}  ~\left ( {L^{(\mu)} \over  \tilde {L} }
\right )^{- {9\over 2}}\, ,
\end{equation}
that is therefore more suppressed at large $L^{(\mu)}$.

\begin{figure} [t]
\centerline{\psfig{figure=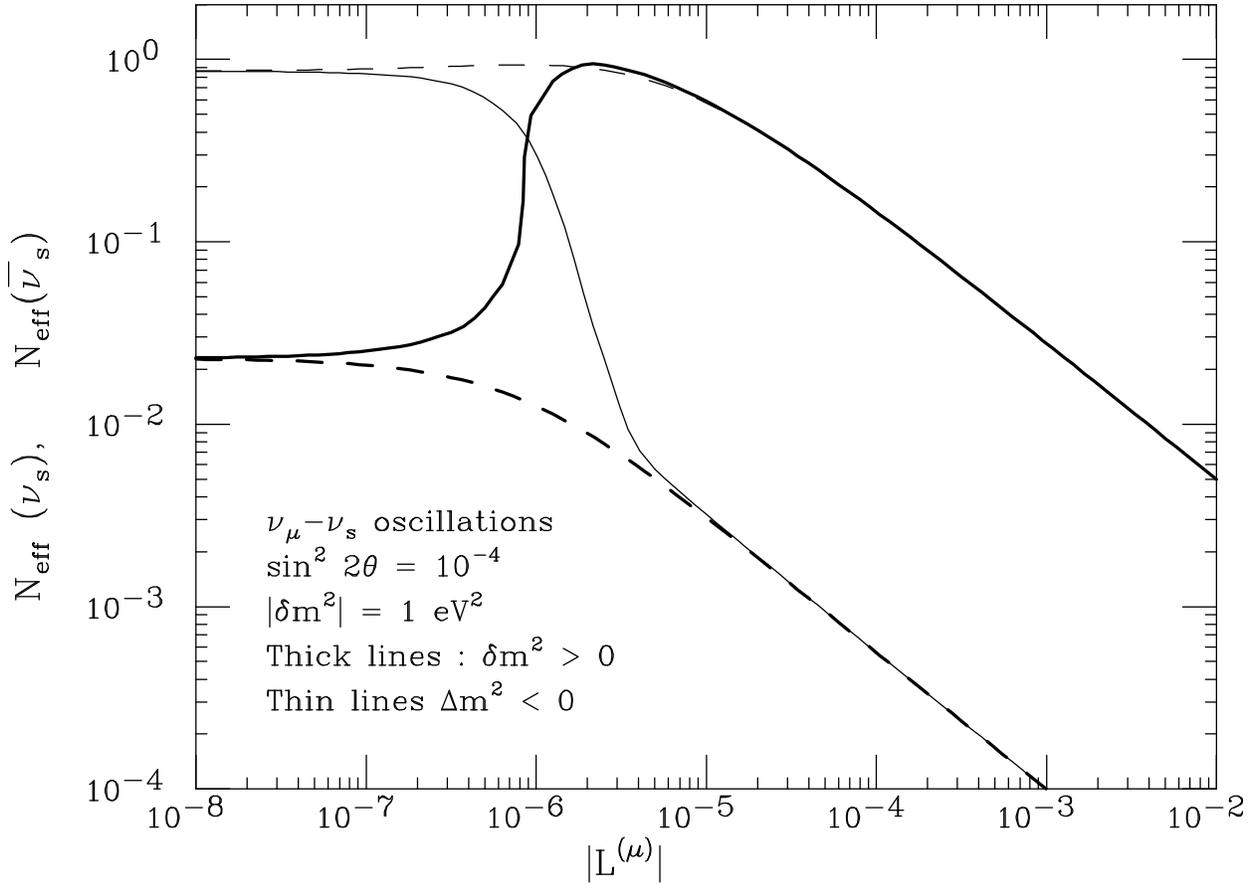,height=14cm}}
\caption {Plot of the effective number   of  sterile neutrinos
$N_{\nu_s}$(eff)    as a function of  $L^{(\mu)}$  
calculated   assuming  that the sterile  neutrino
is   mixed  with a muon
(or  a tau)  neutrino  with   $s^2 = 10^{-4}$ and
$|\delta m^2| = 1$~eV$^2$.
The thick (thin) lines  correspond to 
$\delta m^2 >0$  ($\delta m^2 < 0$).
The solid (dashed) curves correspond to neutrinos 
(antineutrinos)  for  $L^{(\mu)} >0$  and  vice versa for 
$L^{(\mu)} < 0$.
In the  calculation the asymmetry $L^{(\mu)}$ is   fixed
and it is not  considered  as  a  dynamical  variable.
\label{fig:asym2}
}
\end{figure}

\begin{figure} [t]
\centerline{\psfig{figure=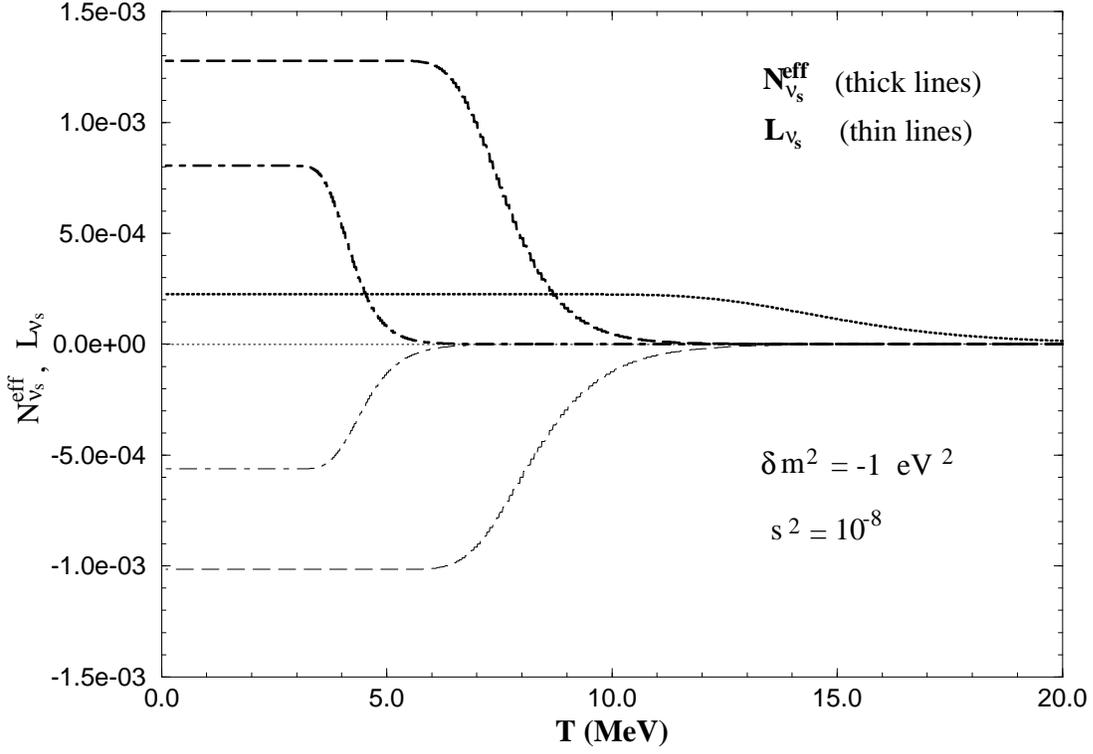,height=16cm,angle=270}}
\caption {
Evolution with temperature of  the density
$N_{\nu_{s}}^{\rm eff}$  and
asymmetry  $L_{\nu_s}$ of  sterile  neutrinos,
calculated 
assuming the existence of  $\nu_s \leftrightarrow \nu_{\mu(\tau)}$ oscillations
with $\delta m^2 = -1$~eV$^2$ and  $s^2 = 10^{-8}$,
and a  time  independent
value for the charge asymmetry of the medium.
The thick (thin) lines  describe  the
evolution  of  $N_{\nu_s}^{\rm eff}$  ($L_{\nu_s}$).
The dotted lines correspond to $L^{(\mu)}= 0$,
the  dashed  to $L^{(\mu)}=10^{-5}$ and the dot--dashed ones to 
$L^{(\mu)}=10^{-4}$.
For $L^{(\mu)} = 0$, 
$\nu$'s and $\overline{\nu}$'s  oscillate in the same  way and 
no  asymmetry is  generated.  For   non  negligible  values
of  $L^{(\mu)}$ the sterile antineutrino production can  be
enhanced,  and  an   asymmetry  is  generated.
\label{fig:dster1}
}
\end{figure}

Note that  for  $s^2 \simeq  1$ the effect of
a  charge asymmetry of the medium   results  in a suppression 
of  the  sterile  neutrino  and  anti--neutrino
population.  The situation   can be different 
 for $s^2$ small.  In this case the presence 
of  a charge asymmetry can even    result in an  enhancement of
the oscillations.
This  is  illustrated in figs.~\ref{fig:asym2} and~\ref{fig:dster1}.
In fig.~\ref{fig:asym2} we 
show a case of enhancement of the sterile production, present 
for $\delta m^2 > 0$  and $s^2 \ll 1$ 
(a similar behavior was also noticed 
for much smaller values of $\delta m^2$ in \cite{Kirilova98}).
For  these oscillation  parameters, in a neutral  medium  there
is  never  a situation of  maximal  mixing,  however for  a 
sufficiently large asymmetry 
neutrinos  or antineutrinos with adimensional  momentum 
$y$  can pass  through a resonance
for two  well defined  values  of the  temperature (see fig. \ref{fig:mix}).
At  the lowest  $T$ value the   production of sterile  neutrinos
(or  anti--neutrinos)  can be  significant, and in fact,
for positive asymmetry, the production of
neutrinos for positive $\delta m^2$ becomes equal to 
the production of antineutrinos for negative $\delta m^2$, since in both
cases we have resonances with almost identical characteristics. 
In fig.~\ref{fig:asym2} the behavior for large 
$ L^{(\mu)} / \tilde {L}$ shows a decrease with the same power law as 
in equation (\ref{eq:Glow}).

For still smaller values of $s^2$, it is possible to have an 
intermediate region
of asymmetries, still larger than $\tilde {L}$, where 
the sterile neutrino production increases with the asymmetry. An 
example of this behavior is given in fig.~\ref{fig:dster1}. In fact, 
integrating  with a saddle  point approximation for  
$ L^{(\mu)} / \tilde {L} \gg 1$, one   finds
for  the resonant  component  (in the approximation $c \simeq 1)$:
\begin{equation}
 G (y, T\to 0)   
\simeq {\pi \over 4} \; \left ( {\tilde{L} \over L^{(\mu)}} \right )^{3 \over 4}
~ \left [ 
 s^2 + d_0^2 ~\left  ( 
{\tilde {L} \over L^{(\mu)}} \right )^{3} \right ]^{-{1\over 2}}
\end{equation}
Note that  for very  large  $L^{(\mu)}$ the  oscillations are  suppressed
as before, however   for a  certain range of   values
of $L^{(\mu)}$  determined by  the value of 
$s^2$ one  can   have  an enhancement. 

It should be clear from the arguments given above that the main
contributions to the integrals come from  rather restricted regions of 
temperature (particularly if a resonance is present) and therefore
the requirement of a constant $L^{(\mu)}$ 
for the validity of these results can be slightly relaxed:
it is in fact sufficient that the asymmetry is slowly varying in the relevant 
region of the integration.

\section{Mixing of two  neutrino flavors}
\label{sec:lae}

 In this section we  will  study the 
  flavor evolution  of    the    neutrino  populations
in the case of  two--flavor mixing  of a standard  and  a sterile
neutrino, considering the   lepton asymmetry of the medium  as  a
dynamical   variable.
We  will  confirm the finding of \cite{Foot96}, that  for 
$\delta m^2 < 0$  one has   in general the generation of
a  large  lepton asymmetry  $L_{\nu_\alpha}$; at the same time,
summing over  $\nu$'s  and $\overline{\nu}$'s, 
the  energy density of the  sterile neutrinos     generated by the 
oscillations   remains similar to what has  been  obtained
in the  previous  section   using the (incorrect)  assumption that
the charge asymmetry of the medium remains    small.

The set of  differential  equations   that  we need  to  solve
(\ref{eq:pbo1}), (\ref{eq:pbo2}), 
(\ref{eq:pbo3}), (\ref{eq:pbo4}) has  been  introduced
in section 2.
In  section 3  we  have  recalled that 
the  collisional operators
maintain  the active neutrinos  in  thermal   equilibrium,  therefore
their  momentum distributions are determined   by  the 
chemical potential $\xi_\alpha (T) = -\bar{\xi}_\alpha (T)$, 
or equivalently  (see equation (\ref{eq:xi})) by the    lepton   number  
$L_{\nu_\alpha}(T)$. Thus the    flavor  evolution is  described
by  a  set of  only two  equations,  
such as (\ref{eq:z})  and (\ref{eq:bz}) 
or  (\ref{eq:zp})  and (\ref{eq:zm}). 
We have  however to take into account   that the  lepton 
number of the medium is  evolving  with time.
The only source of  variation
of  the lepton  number is the presence of oscillations, 
therefore  one has that
$L_{\nu_\alpha} + L_{\nu_s}$  remains constant and:
\begin{equation}
L_{\nu_\alpha} (T)  + {\rm  const} = 
-L_{\nu_s} (T) = 
{1 \over 4\,\zeta(3)}~\int dy ~y^2~[ -f_{\nu_s} (y,T) 
+ f_{\overline{\nu}_s} (y,T)] \; .
\label{eq:asym}
\end{equation}
In summary the  problem  of the flavor evolution
 of  the   neutrino population   has  been reduced
to  the solution  of the  set  of  two  
equations (\ref{eq:z}) and  (\ref{eq:bz}),
with the    additional  condition    that 
the    asymmetry   $L_{\nu_\alpha}$
that  enters  in the expression of the transition rates $\Gamma_{\alpha s}$
and   determines the    distributions  of the   standard
 neutrino and  anti--neutrino population is 
obtained  with the integral (\ref{eq:asym}).

It  is  instructive  to consider 
an equation  directly for  $L_{\nu_{\alpha}}$  that  can  be obtained
combining the definition (\ref{eq:asym})
with  equations  (\ref{eq:z}) and (\ref{eq:bz}). The result is:
\begin{equation}
\label{eq:lnua}
\frac{dL_{\nu_{\alpha}}}{dt}=
{1 \over 4\,\zeta(3)}~\int dy\; y^2~f^{0}_{eq}(y)
~\left[(\Gamma_{\alpha s}-\bar{\Gamma}_{\alpha s})(z^{+}_{s}-z^{+}_{\alpha})+
(\Gamma_{\alpha s}+\bar{\Gamma}_{\alpha s})(z^{-}_{s}-z^{-}_{\alpha})\right] \;.
\end{equation} 
This  equation  allows   to  understand   the evolution of
the    lepton  number  making use of some simple considerations.
The second  term 
tends  to  bring the neutrino  population toward a situation
in which the initial lepton asymmetry is equally shared between
active and sterile neutrinos. The first term, for positive $\delta m^2$,
has always sign opposite to $L^{(\alpha)}$, and also acts as a damping term,
so that the asymmetry approaches a constant value $- \tilde {\eta}/2$, 
with $\tilde {\eta} \equiv L^{(\alpha)} - 2\,L_{\nu_\alpha}$.

On the other hand, if $\delta m^2$ is negative, the first term 
can have either the opposite or the same  sign 
as $L^{(\alpha)}$ and  may also be the source of  a  growth for
the asymmetry.
To  have an appreciable growth we need that
the transition rates  for neutrinos   and  anti--neutrinos  
are  different
and that  the total number of active neutrinos and sterile neutrinos are
also different \cite{Khlopov81}. Therefore
considering   the  oscillation  of two active neutrino
species that have approximately equal number densities,  
one expects only an effect 
of equipartition of the initial asymmetry. A small effect would arise 
after chemical decoupling because of a small number density difference between
the active neutrinos. This is created by electron-positron annihilations 
that reheat only the electron neutrino component \cite{Langacker87}. 
Things can be much more interesting  in the case of active-sterile 
neutrino mixing.

\begin{figure} [t]
\centerline{\psfig{figure=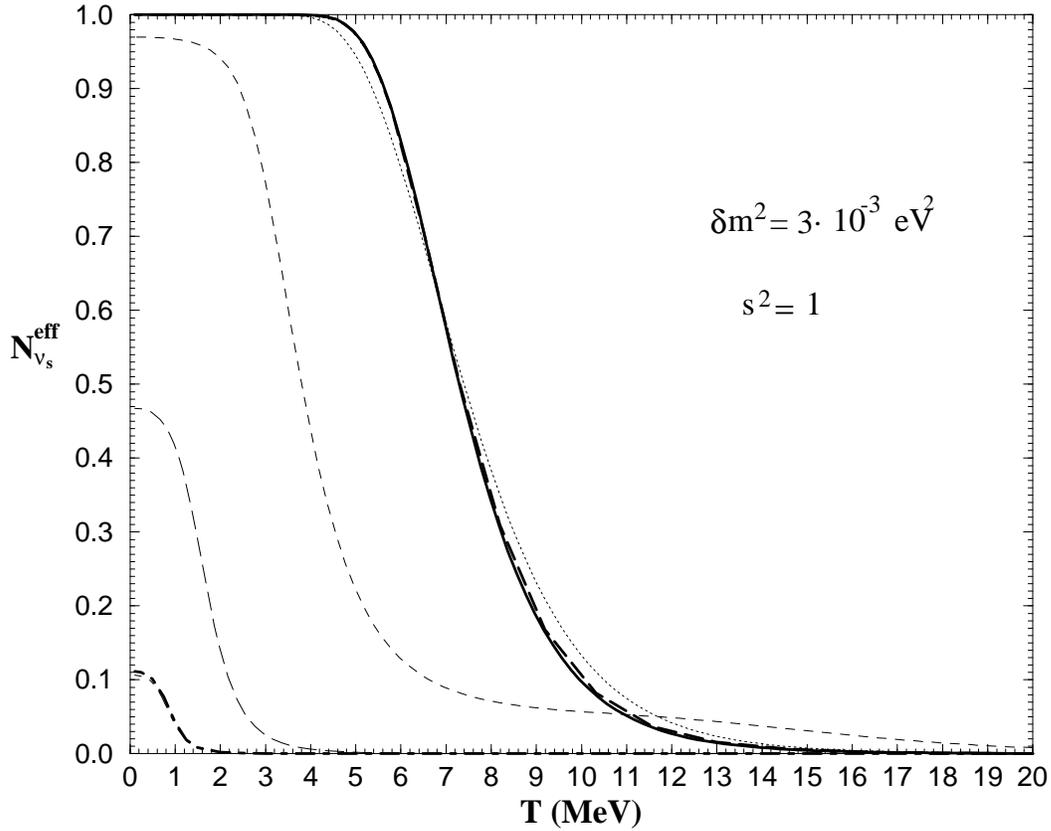,height=16cm,angle=270}}
\caption {
Evolution  of the sterile neutrino energy density
calculated assuming the  existence 
of $\nu_{\mu(\tau)} \leftrightarrow \nu_s$ oscillations
with maximal mixing  and $\delta  m^2 = 3 \times 10^{-3}$~eV$^2$.
 The solid, dotted, dashed,  long dashed   and dot--dashed lines
correspond to an initial  value $L_{in}^{(\alpha)}= 0$,  $10^{-7}$, 
$10^{-6}$, $10^{-5}$  and $10^{-4}$. 
The thin  lines  are  calculated considering $L^{(\alpha)}$ as  a
a  constant, the thick  lines taking into  account the 
dynamical    evolution  of  $L^{(\alpha)}$.
\label{fig:dster}
}
\end{figure}

In  fig.~\ref{fig:dster}
we  show  as  a function of the temperature 
the energy density of  the sterile    neutrinos, always  calculated
assuming the existence 
of $\nu_{\mu(\tau)} \leftrightarrow \nu_s$ oscillations
with maximal mixing, positive $\delta  m^2 = 3 \times 10^{-3}$~eV$^2$ 
and  different initial   values of  the 
asymmetry 
$L^{(\alpha)}$  ($0$,  $10^{-7}$, 
$10^{-6}$, $10^{-5}$  and $10^{-4}$). 
The  thin  curves  show the results  of  a calculation 
performed  considering  
$L^{(\alpha)}$ constant in time: 
increasing $L^{(\alpha)}$ the  sterile  neutrino production is
more and  more suppressed.
The thick lines have  been  calculated
assuming the    initial values    for the asymmetry, 
but taking into  account the 
evolution with time   of the lepton number.
To  a good  approximation   for  any initial  value
$L^{(\alpha)} \lesssim 10^{-4}$ 
the  initial  lepton  number is  efficiently destroyed 
and the evolution of the effective  number of
sterile neutrinos is independent on the initial lepton number 
and  similar to the $L^{(\alpha)} = 0$ case
(three  curves are  coincident with
the thick solid line). 
For  $L^{(\alpha)}\gtrsim  10^{-4}$ the 
lepton number remains 
approximately constant and the results are  very  similar 
to  the case in which the evolution of the lepton number 
with  temperature is  neglected. 

\begin{figure} [t]
\centerline{\psfig{figure=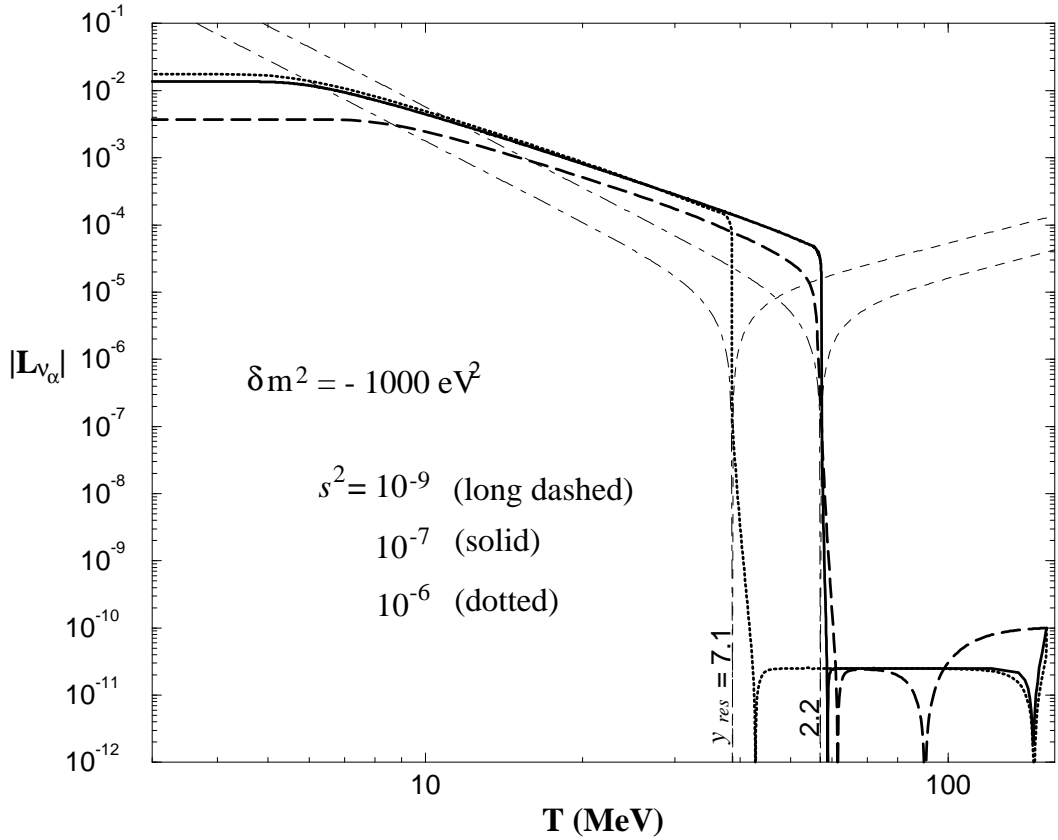,height=16cm,angle=270}}
\caption {
Evolution with temperature of the
lepton number $L_{\nu_\alpha}$  ($\alpha= \mu,\tau$)
(thick lines)  calculated
assuming the existence of 
$\nu_s \leftrightarrow \nu_{\mu(\tau)}$ oscillations 
with  the indicated oscillation parameters.
The   evolution  of the neutrino  populations  is  started 
at $T = 150$~MeV  with $L^{(\alpha)} = 10^{-10}$
and $\tilde {\eta} = 5 \times 10^{-11}$.
The thin lines are the resonance curves for the indicated
values of momenta (dashed for neutrinos and dot-dashed for antineutrinos).
\label{fig:asym-gen}
}
\end{figure}

In fig.~\ref{fig:asym-gen}
we  show  the     
evolution with temperature of the
lepton number $L_{\nu_\alpha}$ 
($\alpha= \mu,\tau$)  calculated   with a    
numerical   integration of the  flavor  evolution equations, 
assuming the existence of 
$\nu_s \leftrightarrow \nu_{\mu(\tau)}$ oscillations 
with a negative $\delta m^2$, and some representative choices
of the mixing  angle. 
The   evolution  of the neutrino  populations  is  started 
at $T = 150$~MeV  with $L^{(\alpha)}_{\rm in} = 10^{-10}$ and 
$\tilde {\eta} = 5 \times 10^{-11}$.

One  can   distinguish  several phases  in  the evolution
of  $L_{\nu_\alpha}$:
(i) initially the    asymmetry $L^{(\alpha)}$ is   erased
and $L_{\nu_\alpha}$ is  
brought to the   value  $ - \tilde{\eta}/2$, 
(ii)  then  at  a  critical  temperature  $T_c$  the asymmetry  
starts  an  exponential  growth,
(iii) after a  short time  the  exponential  growth is transformed  into
a  power  law growth,  and (iv) 
finally the   asymmetry is  `frozen'  and  remains  at a 
constant  value.

Some insight  on   the onset of  the instability can
be obtained   with the following   qualitative discussion, 
where we will assume  for   simplicity that the   oscillating flavors
are $\nu_\tau$ and $\nu_s$.
In this case, as  we have  discussed    in  section~2,  
the resonant  momenta for neutrinos  and anti--neutrinos
are ordered as $y^{\rm res}_{\nu_\tau} > y^{\rm res}_{\overline{\nu}_\tau}$
for  $L^{(\tau)} >0$ 
or 
$y^{\rm res}_{\nu_\tau} < y^{\rm res}_{\overline{\nu}_\tau}$
for  $L^{(\tau)} <0$  (the values are equal in a neutral medium).
Let us  now   consider the evolution of the asymmetry
$L^{(\tau)}$    with  temperature
starting  from an  arbitrary   initial  value.
The values of the resonant  momenta, 
initially very low, increase   with time
$y^{\rm res}_{\nu_\tau} \simeq y^{\rm res}_{\overline{\nu}_\tau}
\simeq \sqrt{c} \,[T^*(\delta m^2)/T)]^{3}$.
When   the resonant  values  become 
of  order  unity  the  oscillations  begin to have  effect
and  start to produce   sterile  neutrinos and anti--neutrinos.
For a  certain   time,
when the position of the  resonant  momenta
is  below  the peak of the Fermi--Dirac   distribution  of 
the   standard neutrinos,
no lepton  asymmetry is   produced
and  in fact  any small  initial  non--zero   value 
of $L^{(\tau)}$ is erased.
This  can be  easily
understood   observing that  if 
in this  situation
(when both    resonant  values  of the  adimensional 
momentum are   below  the peak of the  distribution)
a small  positive (negative) $L^{(\tau)}$  develops,
the  resonance for the    neutrinos  will move to
higher  (lower) values of $y$, with respect
to  the  position of the resonance for  anti--neutrinos;
therefore  the   neutrinos   will oscillate
more (less) efficiently, and 
since  the sterile   particles
are much  less  abundant that  the 
standard  ones, this  will 
result  in  the  net disappearance of $\nu_\tau$'s   
($\overline{\nu}_\tau$'s)
and  in  the generation of a    negative (positive) $L_{\nu_\tau}$ that 
will tend to cancel the initial fluctuation of the 
charge number $L^{(\tau)}$.
In summary  
an initial value  of the charge
asymmetry  will    be  erased 
and $L^{(\tau)}$  will   take  a  value close to  zero.
This  mechanism  of    destruction of an
initial  charge asymmetry   cannot  work if  the  initial asymmetry is  too
large  and  becomes  inefficient  for $|L^{(\alpha)}| \gtrsim 10^{-4}$
\cite{Foot95}.

The   argument  outlined   above 
to  justify the stability of  the  condition
$L^{(\tau)} = 0$  obviously loses its  validity when    the resonant  momenta
are   above the peak of the Fermi--Dirac distribution, 
at $y \simeq 2.2$.
More precisely when the  temperature  drops  below a critical  value
$T =  T_c$ such  that 
\begin{equation}
y^{\rm res}_{\nu_\tau} \simeq y^{\rm res}_{\overline{\nu}_\tau} 
\simeq \sqrt{c}\, \left(  {T^*\over T_c } \right)^3 \simeq  2.2
\end{equation}
the situation  becomes  unstable.  A small   positive  fluctuation of 
$L^{(\tau)}$  results in
a  larger resonant  value    for  the neutrinos  
so that   the    oscillations of  the anti--neutrinos
become more  efficient, more  $\overline{\nu}_\tau$'s  disappear
and the asymmetry $L_{\nu_\tau}$  increases even more.
The same  argument   shows  that an initial  negative
fluctuation in $L^{(\tau)}$  after  $T_c$, 
would grow  to  a much  larger 
(in absolute value) negative  asymmetry.

At first sight   it  appears  that  we have a genuine 
case of   instability  and  that   the   sign of the 
lepton number will be determined  only by random   fluctuations.
This is  in   fact {\em not} the case, at least in the
theoretical  framework we
are considering.  It is  possible  to    demonstrate 
that asymptotically  (for $T \to 0$) the sign of the  lepton asymmetry
 will be the same  as the sign of 
$\tilde {\eta} = L_{\nu_e} + L_{\nu_\mu} - B_n/2$, 
which is the 
charge   asymmetry of the medium   excluding the contributions
of the $\nu_\tau$'s and 
 that  in the  present  discussion remains constant \cite{Foot97}.
Note that it can   be expected  
that $\tilde  {\eta}$ is  positive.
This   can be derived  from the assumptions  that
(i)   the electric  charge of the universe  vanishes
($L_e = B_p$),
(ii)  the net quantum number $B - L$ of the universe  also 
vanishes,
(iii)   protons  and  neutrons  have equal number densities (for $T > 3$~MeV)
and therefore $B_p = B_n$,
and (iv)    the total lepton number of  the standard  neutrinos 
before the development of the oscillations
is equally divided  among all flavors
($L_{\nu_e} = L_{\nu_\mu} = L_{\nu_\tau}$).
From  these  assumptions one  can  derive
$\tilde {\eta} = B_p/6$, that is  a positive  number
of order $10^{-10}$.
At the critical temperature, where the term proportional to 
$L_{\nu_\tau}$ in the right hand side of equation (\ref{eq:lnua}) vanishes, and
the sterile neutrino asymmetry is still negligible, the sign of the
time derivative of $L_{\nu_\tau}$ is the same as the sign of $\tilde {\eta}$;
therefore, once the term driving the instability starts to dominate, 
$|L_{\nu_\tau}|$ will continue to grow, and the sign will be the same as 
the sign of $\tilde {\eta}$.
We will come back to this point and briefly reconsider the limitations of
the present approach in the concuding section.

\begin{figure} [t]
\centerline{\psfig{figure=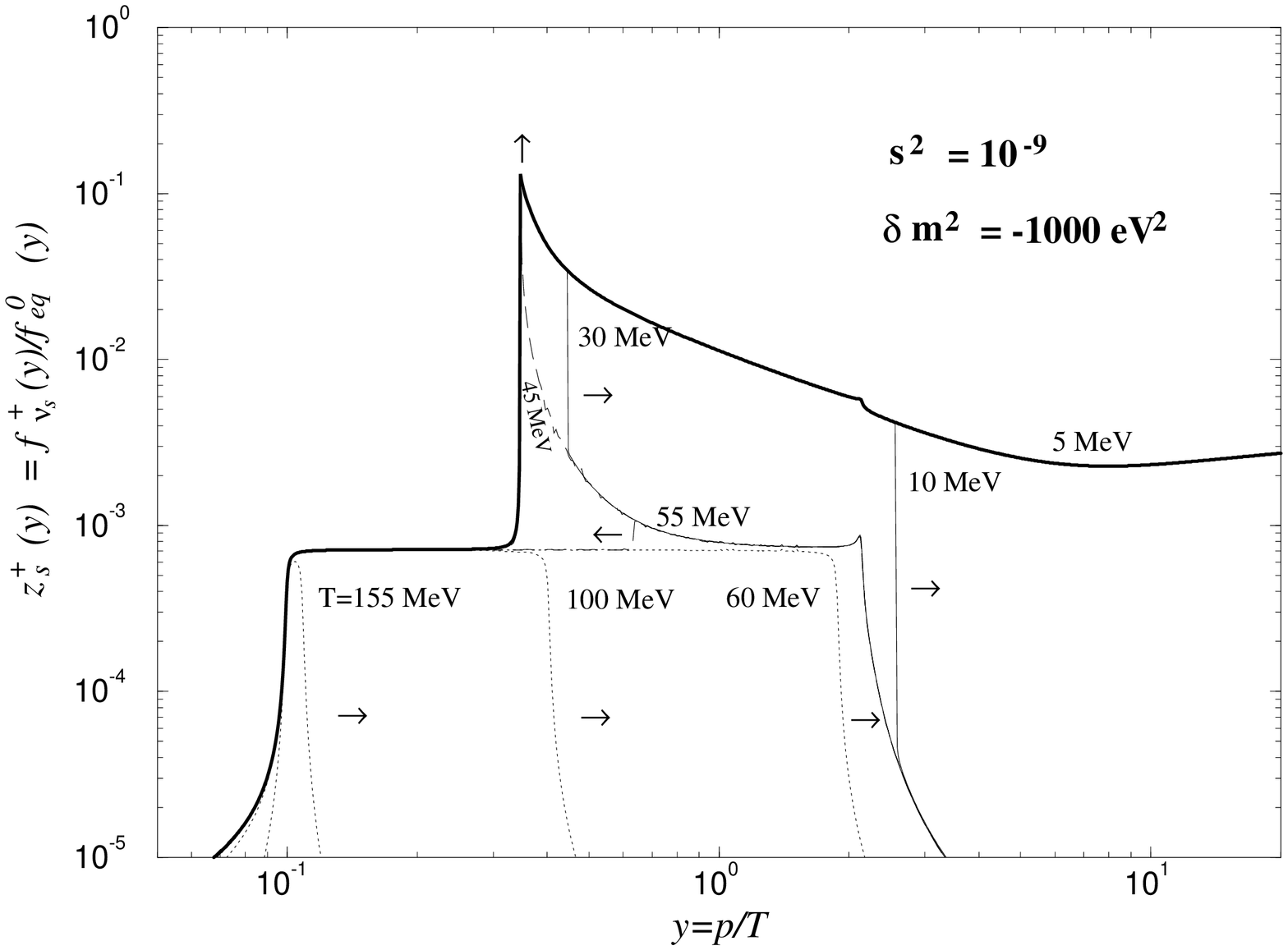,height=16cm,angle=270}}
\caption {
Evolution of the sterile neutrino distribution with
temperature for $\nu_s \leftrightarrow \nu_{\mu(\tau)}$  oscillations 
when lepton number is generated. 
The oscillation  parameters  are $\delta m^2 = -1000$~eV$^2$ and
$s^2 = 10^{-9}$ and correspond to the thick long-dashed line in figure 
\ref{fig:asym-gen}. 
The  different  curves  correspond  to   increasing  temperature.
Above the critical temperature ($T_c \sim 60$~MeV) a standard production
with zero lepton number occurs, up to $y^{\rm res}\simeq 2.2$. 
At lower temperature, 
an enhancement regime with constant $y^{\rm res}\simeq 0.35$ is clearly 
apparent. 
The production occurring after lepton number generation gives a dominant 
contribution in this case, but the total production is anyway too small to 
be relevant for BBN bounds.
\label{fig:evol1}
}
\end{figure}

\begin{figure} [t]
\centerline{\psfig{figure=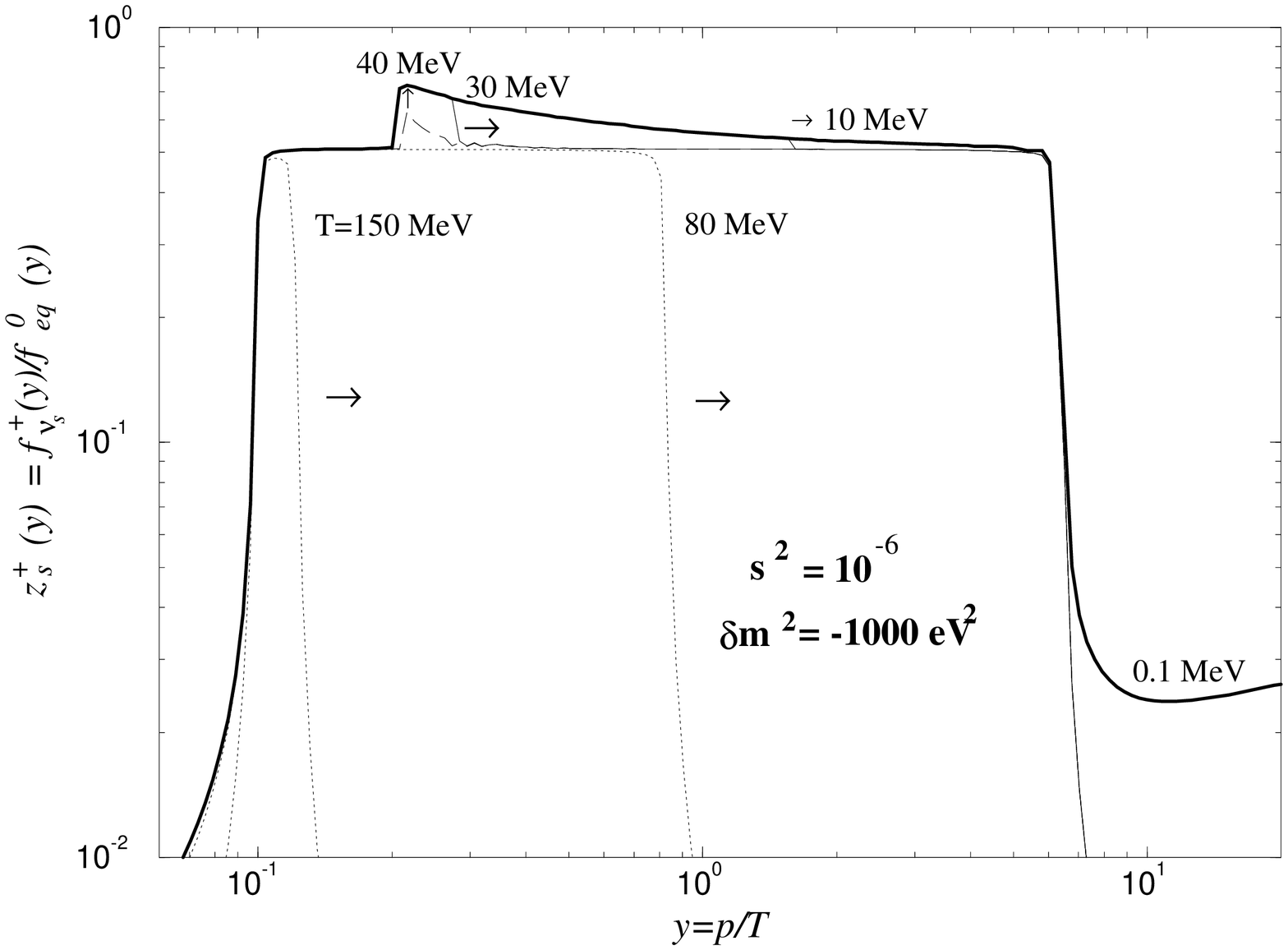,height=16cm,angle=270}}
\caption {
Evolution of the sterile neutrino distribution with
temperature for $\nu_s \leftrightarrow \nu_{\mu(\tau)}$  oscillations 
when lepton number is generated. 
The oscillation  parameters  are $\delta m^2 = -1000$~eV$^2$ and
$s^2 = 10^{-6}$ and correspond to the thick dotted line in figure 
\ref{fig:asym-gen}. 
The  different  curves  correspond  to   increasing  temperature.
Above the critical temperature  ($T_c \sim 42$~MeV) a standard production
with zero lepton number occurs. It can be noticed that in this case 
$y^{\rm res} \simeq 7.1$, since $z^{+}_{s}\lesssim 1$. 
The enhancement regime now is too short to be observed and the dominant 
contribution to the total production comes from the standard regime with zero 
lepton number.
\label{fig:evol2}
}
\end{figure}

In fig.~\ref{fig:evol1}
and~\ref{fig:evol2}
we show the evolution of  the  distributions
of   sterile  neutrinos  and  antineutrinos
with temperature.   
We can  again  distinguish  different    regimes.
In  a first phase, for 
 $T$  larger  than  a  critical   temperature  $T_c$,
the    initial small  asymmetry  of the medium is  erased, 
the     resonant  momentum for  neutrinos  and  anti--neutrinos
grows  ($y^{\rm res} \propto T^{-3}$) producing  sterile  neutrinos
with higher  and  higher  momentum.
During  this phase  (as  discussed in  section 4) 
the  sterile neutrino population,  for $y <y^{\rm res}(T)$,
is proportional to  a  thermal  distribution,  that is
$z_s(y)$  and $\bar{z}_s(y)$ are  constants up to   a  maximum
$y  \lesssim y^{\rm res}(T)$
(compare with  fig.~\ref{fig:evol0}).   

When  the temperature drops  below
the  critical  temperature  and the production of 
a positive $L_{\nu_\tau}$ starts, the  resonant   momentum
for  anti--neutrino  oscillations is  pushed back to  lower   values 
producing   additional   sterile  anti--neutrinos.
In fig.~\ref{fig:evol1} it appears clearly that  
the critical  temperature corresponds  to  the situation where the
resonant  momentum  was  $y^{\rm res}_{\nu_\tau(\overline{\nu}_{\tau})}
 \simeq 2.2$.

Fig.~\ref{fig:evol2}   is  calculated  for the same
value  of $\delta m^2$, but for  a larger  mixing parameter
$s^2$.  In this case the 
critical  temperature  is  lower  and  corresponds to
a  situation  where the
resonant  momenta  for  neutrinos  and anti--neutrinos
is larger,  $y^{\rm res}_{\nu_\tau(\overline{\nu}_\tau)} \simeq 7.1$.
An understanding of this ``delay'' of the onset of the instability
for large $s^2$ can be obtained inspecting  equation
(\ref{eq:lnua}).  It can  be seen  that  the  term  that  controls  the
growth of the  asymmetry is 
$(\Gamma_{\alpha s} - \bar{\Gamma}_{\alpha s}) (z_s^+ - z_\alpha^+)$.
This  term   is   reduced  if the sterile  neutrinos    develop
a  sizeable  population, as  it is the case  for a  sufficiently large
$s^2$. We  can  see in fig.~\ref{fig:evol2}  that the  
sterile  neutrinos  produced
in the early phase ($T > T_c$) are  in fact  a  significant  fraction of
the   thermal  equilibrium  population.

The   estimate  of the 
energy  density in  sterile neutrinos  and anti--neutrinos
produced  by oscillations  for  $T \to 0$,
 is  modified   with respect to
the estimates  developed in the previous  section,  because
of the effects of the  evolution of the lepton asymmetry
that   determines  the neutrino  effective potential.
It is  important to note  that the  difference between the  estimates
(with and  without the  dynamical  evolution of $L_{\nu_\alpha}$)
is  small   for those   oscillation parameters
that    predict  an appreciable 
energy density, $N_{\nu_s}^{\rm eff} \gtrsim 0.3$.
This is  illustrated in fig.~\ref{fig:final}  where the  two   thin
solid lines  in  the plane 
($s^2$, $\delta m^2$) correspond
to $N_{\nu_s}^{\rm eff} = 0.1$  and~0.6  in a  complete  calculation
of the flavor  evolution, 
while the  two thin  dot--dashed  lines  are 
calculated assuming  a constant negligible $L_{\nu_\alpha}$.
Note  that there is  some difference in the results
for $N_{\nu_s}^{\rm eff} = 0.1$ (so that in the complete calculation
the region having $N_{\nu_s}^{\rm eff} < 0.1$ is larger), but 
the two lines corresponding to 
$N_{\nu_s}^{\rm eff} = 0.6$  are  essentially undistinguishable.
This can  be understood  qualitatively (see also  the discussion in
\cite{Foot97,Foot97c}):
when the production of sterile   neutrino is   sufficiently strong,
the   lepton number  generation is
delayed (the  critical   temperature  becomes  lower), and 
the subsequent production of sterile  neutrinos
in the regime with a not negligible lepton number
occurs far from the peak of the Fermi distribution, and therefore gives  
a small  contribution, see fig.~\ref{fig:evol2}.

Thus we can conclude that in the simplest  framework of two--flavor
active-sterile  neutrino mixing 
a  detailed   treatment of  the lepton  number  generation 
does not essentially modify the region  of oscillation parameters that 
the  constraints of  BBN  can exclude. 
As  we  will   discuss in the next section, this is not  true any more
if we consider a more   general  framework with 
more elaborate forms  of  mixing  between   neutrino  flavors.

\section{Three  neutrinos  mixing scenario}
\label{sec:3nu}

In this  section we  will consider   the 
scenario proposed  by Foot  and Volkas \cite{Foot97}
where  a  sterile  neutrino is  mixed  with   both 
$\nu_{\mu}$ and $\nu_{\tau}$. 
As in \cite{Foot97}  we  will not  consider  the most   general 
case  of  three  neutrino mixing
 but we  will  limit  our  considerations  to a situation 
where   we can  neglect  the $\nu_\mu \leftrightarrow  \nu_\tau$ oscillations
and  the descriptions  of all  transitions  can be well  
approximated  
by  a pair  of two--flavor  oscillations:
$\nu_{\mu} \leftrightarrow \nu_{s}$  and $\nu_{\tau} \leftrightarrow \nu_{s}$.
In the scenario we are considering
two  neutrino  eigenstates  are  nearly  degenerate
with a squared mass  difference of order  
$\delta m^2_\mu \sim 3 \times 10^{-3}$~eV$^2$, and are  both   superpositions
with approximately equal  weights  of $\nu_\mu$  and $\nu_s$
and  only a small  component of $\nu_\tau$; 
in this way  the $\nu_{\mu} \leftrightarrow \nu_{s}$   oscillations
can  describe the atmospheric  neutrino data. 
The third mass  eigenstate  is nearly a pure $\nu_\tau$. 

As we  have seen before, if the 
$\nu_\tau$ does  not  participate  in the oscillations,
the  $\nu_\mu \leftrightarrow \nu_s$   transitions
result in an  energy density in sterile  neutrinos
not compatible with suggested bounds from big bang nucleosynthesis,
unless an unnaturally  high lepton number $L^{(\mu)}$ is 
present at the start and persists down to temperatures $T \lesssim 10$~MeV.
The inclusion of 
a small mixing  of the $\tau$ neutrinos   deeply modifies this
conclusion.
The oscillations 
between  $\nu_{\tau}$ and   $\nu_{s}$   can   produce
a  significant   $L_{\nu_\tau}$  and   this in turn 
can  suppress the 
production of sterile neutrinos in the  $\nu_{\mu} \leftrightarrow \nu_{s}$ 
oscillations.

For  this  mechanism  to   work 
it is necessary to require 
$|\delta m^{2}_{\tau}| \ge |\delta m^2_\mu| $ (so that
the  $\nu_{\tau} \leftrightarrow \nu_{s}$ 
oscillations  start  earlier),
$\delta m^{2}_{\tau}<0$   (the 
mass   eigenstate   dominantly coupled  to $\nu_\tau$ is
the  heaviest) and 
$s^{2}_{\tau}$ sufficiently small 
(to assure that the  lepton number generation can actually occur). 
Moreover the constraint (\ref{eq:constr}) must be verified, otherwise  
a sterile neutrino overproduction would occur already during the early 
$\nu_{\tau} \leftrightarrow \nu_{s}$ oscillations. These conditions are 
not sufficient by
themselves to prevent a subsequent destruction of the lepton number by the
$\nu_{\mu} \leftrightarrow \nu_{s}$ oscillations and as a consequence
$N^{\rm eff}_{\nu_s} \simeq 1$. In the following the
conditions on the oscillation parameters to avoid this conclusion 
are discussed.

We indicate with $s_{\alpha}$, $c_{\alpha}$, 
$\delta m^{2}_{\alpha}$
($\alpha=\mu,\tau$) the mixing parameters for the 
$\nu_{\alpha} \leftrightarrow \nu_{s}$  oscillations.
The set of equations that  describe  the  flavor  evolution
for the proposed scenario 
can   be easily obtained  with  a simple  generalization of 
the  case  of  two  flavor  mixing discussed in section \ref{sec:lae}, 
describing at the same time the dynamical evolution of the two lepton numbers 
$L_{\nu_{\mu}}$ and $L_{\nu_{\tau}}$. 
As before the active  (anti--)neutrinos  are  assumed  to be 
in thermal equilibrium, namely, neglecting terms of second order in
the  chemical potential, 
  $z_\mu^+  = z_\tau^+ \simeq  1$, $z_{\mu(\tau)}^- \simeq 
12\;\zeta(3)\;L_{\nu_{\mu(\tau)}}/[\pi^2\;(1+e^{-y})]$.
Thus we can write:
\begin{equation}
{dL_{\nu_\tau} \over dt}=
-{1 \over 4\,\zeta(3)}~\int dy\;y^2 ~f^0_{eq} (y) \left \{
[\Gamma_{\tau s}  - \overline{\Gamma}_{\tau s}] \;
\left ( 1 - z_s^+ \right )
+
[\Gamma_{\tau s}  + \overline{\Gamma}_{\tau s}]\,
\left (z_\tau^- - z_s^- \right )  \right \}
\label{eq:ACCtau}
\end{equation}
\begin{equation}
{dL_{\nu_\mu} \over dt}=
-{1 \over 4\,\zeta(3)}~\int dy\;y^2 ~f^0_{eq} (y) \left \{
[\Gamma_{\mu s}  - \overline{\Gamma}_{\mu s}] \;
\left ( 1 - z_s^+ \right )
+
[\Gamma_{\mu s}  + \overline{\Gamma}_{\mu s}]\,
\left (z_\mu^- - z_s^- \right )  \right \}
\label{eq:ACCmu}
\end{equation}
\begin{equation}
\frac{d}{dt}z_{s}(y,t)=
[\Gamma_{\tau s}+\Gamma_{\mu s}]\left[1-z_{s}(y,t)\right]
\label{eq:ACCst}
\end{equation} 
\begin{equation}
\frac{d}{dt}\bar{z}_{s}(y,t)=
[\overline{\Gamma}_{\tau s}+\overline{\Gamma}_{\mu s}]
\left[1-\overline{z}_{s}(y,t)\right]
\label{eq:ACCbst}
\end{equation}
Note that  the   evolutions  of the two asymmetries  $L_{\nu_\mu}$
and $L_{\nu_\tau}$ are  coupled   for  two  reasons, the   first being 
simply  that  $\nu_s$ 
oscillate  both in  $\nu_\mu$ and in $\nu_\tau$, the second is that,
also  when the  population of sterile neutrinos
is small and  the  transitions  $\nu_s \to \nu_{\mu(\tau)}$ can be
neglected,  the   asymmetry  of one type of flavor
(for  example  $L_{\nu_\tau}$)    is  relevant 
for the  evolution  of  the other  asymmetry  
(in this  case $L_{\nu_\mu}$)   because it modifies
the  effective  potential in the medium of the neutrino ($\nu_\mu$).

We have  numerically  integrated  this  set of  equations
calculating as a function of temperature  the     distributions
of sterile  neutrinos  and  the asymmetries of 
$\nu_\mu$'s  and $\nu_\tau$'s.   The results  show  some
very  interesting   and  not entirely intuitive  features.
In fig.~\ref{fig:final}   we show   curves  of  constant  energy density
in  sterile neutrinos  for  $T \to 0$  as  a function of
the oscillation   parameters $\delta m^2_\tau$  and $s^2_\tau$
assuming $s^2_\mu = 1$ and a fixed  value  of
$\delta m^2_\mu$.
The most  remarkable  feature is  that for
a given   value of  $\delta m^2_\mu$    the      plane
($s^2_\tau$, $\delta m^2_\tau$) is  divided  in  two  regions.
In one (the region  below the thick  lines) 
the  sterile neutrinos become  fully thermalized, while in the other
(above  these  lines) the  energy density in sterile neutrinos
drops  to  a   much  lower  value. This value is almost undistinguishable
from what is  obtained  in a  calculation with only 
$\nu_\tau \leftrightarrow  \nu_s$ oscillations (the 
$\nu_\mu\leftrightarrow  \nu_s$ oscillations being completely 
damped).
In fact, the lines of constant $N_{\nu_s}^{\rm eff}$ in
fig.~\ref{fig:final} drawn as thin lines are exactly the same that
were obtained and discussed in section \ref{sec:lae} using a 
two--neutrino scenario.

The   boundary   in parameter  space between the region 
where  the sterile  neutrino is thermalized 
(resulting in $N_{\nu_s}^{\rm eff} \simeq 1$) 
and the region  where  the effect of  
$\nu_\mu \leftrightarrow \nu_s$ oscillations 
is  essentially negligible  is   given  to  a  first approximation
by the relation
\begin{equation}
s^2_\tau \; (\delta m^2_\tau)^2 \simeq \; 28 \; (\delta m^2_\mu)^2 
\label{eq:approx}
\end{equation}
(see the dashed   curves  in  fig.~\ref{fig:final}). From the figure,
we read that the minimum value allowed for the tau--neutrino mass in 
order to comply the BBN bounds is $\sim 1.4$~eV (3~eV) if 
$\delta m^2_\mu = 10^{-3}\ (3.2 \times 10^{-3})$~eV$^2$, with a mixing 
angle $s^2_\tau \simeq 10^{-5}$.

In fig.~\ref{fig:boundary}  we show the   evolution  with temperature
of the  sterile  neutrino effective  number  
considering values for the neutrino  oscillation  parameters  that
are very close to  each other  but belong to the two
different  regions where the sterile neutrinos are or are not 
thermalized. The thick lines correspond to a point immediately below
the corresponding boundary in fig.~\ref{fig:final}: the sterile neutrino 
production is momentarily interrupted when the $\nu_\tau$ asymmetry is
generated at a temperature of about 30 MeV, but, since 
the initial rapid increase of $L^{(\mu)}$ is stopped and then reversed,
it starts again and reaches 
the thermal equilibrium at a temperature $\sim 10$~MeV. The thin lines, 
corresponding to a point immediately above 
the boundary in fig.~\ref{fig:final}, 
describe a situation in which $L^{(\mu)}$ continues to increase, and the 
$\nu_\mu\leftrightarrow  \nu_s$ oscillations are damped and unable to
generate further sterile neutrinos.

In the following  we will present some  
qualitative arguments and analytic estimates 
to illustrate these numerical results.

\subsection {Qualitative  discussion}
In fig.~\ref{fig:film}  we  show  an  illustration of 
the   neutrino mixing parameters  in matter 
at    different  times    (or  temperatures).
This particular    example  was  calculated  for the set
of  oscillation   parameters:
$\delta m^2_\mu = 10^{-3}$~eV$^2$, 
$s^2_\mu = 1$, $\delta m^2_\tau = -10$~eV$^2$, 
$s^2_\tau = 10^{-6}$.
In each panel the four  different curves describe the effective
mixing parameter for the  four transitions
between  standard  and  sterile  (anti--)neutrinos:
$s^2_{\mu,{\rm m}}$,
$\overline{s}^2_{\mu,{\rm m}}$,
$s^2_{\tau,{\rm m}}$,
and $\overline{s}^2_{\tau,{\rm m}}$;
 the different  panels  refer  to  different    temperatures, 
from top  to  bottom the temperatures and asymmetries 
in the  five  panels correspond to  the situations
(a) : $T= 58.1$~MeV,  $L_{\nu_\tau} = 0$,
(b) : $T= 38.7$~MeV,  $L_{\nu_\tau} = 0$,
(c) : $T= 34.9$~MeV,  $L_{\nu_\tau} =  8.9 \times 10^{-7}$,
(d) : $T= 32.2$~MeV,  $L_{\nu_\tau} =  1.7 \times 10^{-6}$,
(e) : $T= 29.1$~MeV,  $L_{\nu_\tau} =  3.2 \times 10^{-6}$
and in all cases $L_{\nu_\mu} \simeq 0$.
The  behaviour of the  
evolution of the  mixing parameters  is  easy to  understand
qualitatively.
At  high  temperatures  the  medium is   neutral  and   the
oscillation  parameters in matter  for $\nu$'s  and $\overline{\nu}$'s are
equal.   
For   vanishing  charge  asymmetry of the medium,
the mixing parameter  in matter for $\nu_\mu \leftrightarrow  \nu_s$ 
oscillations is
\begin{equation}
 s^2_{\mu,{\rm m}}  =   \overline {s}^2_{\mu,{\rm m}}  = 
\left [ 1 +  \left ( {b_\mu  \, T^6 \,y^2 \over \delta m^2_\mu  } \right )^2
\right ]^{-1} 
\simeq  \left ( {T^*_\mu \over T} \right )^{12} \; y^{-4}
\end{equation}
and  the  mixing is  strongly suppressed  by  matter effects.
For the $\nu_\tau \leftrightarrow  \nu_s$ 
oscillations,  the   important   feature is the 
fact that  at a   momentum 
\begin{equation}
y^{\rm res}_{\nu_\tau} = y^{\rm res}_{\overline{\nu}_\tau} = \sqrt{c_\tau}
  \left ( { \delta m^2_\tau \over b_\tau}
\right )^{1\over 2}   \; T^{-3} = \sqrt{c_\tau}
 \left ( {T^*_\tau \over T} \right)^3
\end{equation}
the  mixing  for  neutrinos  and  antineutrinos  is  maximal.
With decreasing  temperature     the  suppression 
of the   $\nu_\mu \leftrightarrow  \nu_s$ 
oscillations becomes   weaker 
while  the 
position of the   resonance   for the $\nu_\tau \leftrightarrow \nu_s$ 
oscillations grows  as $T^{-3}$.
When  $y^{\rm res}_{\nu_\tau}$  approaches the  peak of the Fermi distribution, 
the  instability  discussed  previously sets in
and an asymmetry $L_{\nu_\tau}$  becomes  to develop.
The  positions  of the   resonances  for the
tau-sterile oscillations  of neutrinos  and  anti--neutrinos are 
given by equation (\ref{eq:yres_}) for $\alpha$=$\tau$.
If a positive  $L_{\nu_\tau}$  starts  to  be generated
the resonance for the  neutrinos  moves to higher  momenta $y$,
where the neutrino   population is  smaller,
therefore  $\overline{\nu}_\tau$  oscillate  more  effectively,  and
the asymmetry  grows   faster.

An  important consequence of the 
generated  $L_{\nu_\tau}$ is  to  modify
the  oscillation  parameters
of the $\nu_\mu \leftrightarrow  \nu_s$  oscillations.
The presence  of  an  $L^{(\mu)}$   asymmetry 
depresses  the oscillations
for  most    values  of the neutrino  momentum, but also
induces   maximal  mixing for  neutrinos   or  antineutrinos
(depending on the sign  of $L^{(\mu)}$)
in a  narrow    range  of momenta.
For  positive 
$L^{(\mu)} \simeq L_{\nu_\tau}$ the
resonance  is present for neutrinos at a  momentum
\begin {equation}\label{y_res}
y^{\rm res}_{\nu_\mu}  = {a_L \over b_\mu} \, {L^{(\mu)} \over T^2} 
\end{equation}
and it gives rise to the  generation  of   a  
negative  $L_{\nu_\mu}$,  that   also acts   on the   oscillation
parameters   for both type of  oscillations.

It   appears    clearly   now  that
two  different  scenarios  can  develop.
In  the  first case the  generation  of 
 $L_{\nu_\mu}$   is  never   sufficiently rapid  to    
cancel the growth of   $L_{\nu_\tau}$, 
and the   combined  asymmetry
$L^{(\mu)} \simeq 2 L_{\nu_\mu} + L_{\nu_\tau}$ keeps    growing.
The   resonant  peak for the  muon neutrino oscillations
rapidly  ($y^{\rm res}_{\nu_\mu} \propto  L^{(\mu)} /T^2$) 
 moves  to very  high values of  $y$  where  it  becomes  irrelevant
because no neutrinos  have such a high momentum,
and the $\nu_\mu \leftrightarrow  \nu_s$ 
oscillations remain suppressed by the generated $L^{(\mu)}$ 
(this is the situation envisaged in fig. \ref{fig:film}).
In the second  possible  outcome,
the   oscillations $\nu_\mu \leftrightarrow  \nu_s$  proceed so rapidly
that at  a certain  time 
the  growth  of  $L^{(\mu)}$ is  stopped  and  then reversed.
Qualitatively it is  understandable that in the space of
oscillation  parameters  there will be a  sharp boundary 
between  regions  where  the two  different scenarios develop.
For example let us keep fixed
$\delta m^2_\mu$, $\delta m^2_\tau$ and $s_\mu^2 = 1$,
and consider the evolution of  the  neutrino populations
for  different   $s^2_\tau$. 
In a certain  interval  of $s^2_\tau$ the oscillations of tau--neutrinos 
proceed   sufficiently  rapidly  
to generate a  net $L^{(\mu)}$  and  suppress the 
muon--neutrino oscillations.
Decreasing  $s^2_\tau$ the rates for the $\nu_\tau \leftrightarrow \nu_s$
transitions, equations (\ref{eq:gamma}, \ref{eq:bgamma}), 
become weaker and weaker
and  the  generation of  $L_{\nu_\tau}$  slower.
At  a  critical   value  the $\overline{\nu}_\tau \leftrightarrow 
\overline{\nu}_s$    
oscillations do  not  proceed   sufficiently rapidly to  cancel
the effect  of the  $\nu_\mu \leftrightarrow \nu_s$ oscillations
and the growth of $L^{(\mu)}$ can  be stopped and reversed.

Qualitatively it should be clear that 
for  a set of  parameters  along the  boundary between the two  regions
the   conditions   (temperature  and  asymmetries  of the medium)
when  the derivative  $dL^{(\mu)}/dt$  vanishes and the growth
of the asymmetry is  reversed    are such that the 
resonance for the $\nu_\mu \to \nu_s$  transitions  is close
to the maximum of the Fermi distribution
($y^{\rm res}_{\nu_\mu} \simeq  y_{\rm peak} = 2.2$), where it affects   
the largest  possible neutrino population.

\subsection {Analytic  estimate}
To obtain  an estimate of the 
boundary  that separates the regions in the space
of the neutrino  oscillation parameters  
where  the sterile neutrino production 
is suppressed, we   will make  the assumption  that  the
reversal  of the asymmetry $L^{(\mu)}$ happens
when  both the   resonances for
$\nu_\mu \leftrightarrow \nu_s$ 
and $\overline{\nu}_\tau \leftrightarrow \overline{\nu}_s$  
oscillations, that    are the source
of a  negative $dL_{\nu_\mu}$  and
a  positive 
$dL_{\nu_\tau}$  
are  most effective,  that  is  in  conditions  where 
both  resonances  are 
at a value  $y \simeq y_{\rm peak} \simeq 2.2$
(this  corresponds to the situation of panel (d)  in fig.~\ref{fig:film}).
We will therefore impose the condition:
\begin{equation}
 \left [ 2 {dL_{\nu_\mu} \over dt} + {dL_{\nu_\tau} \over dt} 
\right ]_{y^{\rm res}_{\nu_\mu} = y_{\overline{\nu}_\tau}^{\rm res}
 = y_{\rm peak}}  = 0
\label{eq:limit2}
\end{equation}
This    algorithm  is  motivated  by  our    numerical results  that 
indicate   that the  reversal  happens
for    values of the temperature  and the asymmetries $L_{\nu_\mu}$
and  $L_{\nu_\tau}$  such  that this  condition
is  approximately satisfied.
The  algorithm  can  also be  understood  qualitatively 
observing that  if there is  no  reversal of
the  asymmetry $L^{(\mu)}$   the  resonance  momentum 
$y^{\rm res}_{\nu_\mu}$    takes all  values from zero to  infinity  
and therefore it will take the value  $y_{\rm peak}$; the  effect of the
$\nu_\mu \leftrightarrow \nu_s$ oscillations  (relative to the
$\overline{\nu}_\tau \leftrightarrow \overline{\nu}_s$ oscillations) is  
on the other hand most important when $y^{\rm res}_{\nu_\mu} 
\simeq  y_{\rm  peak}$.
During the early phase of the generation  of $L_{\nu_\tau}$
one also  has   $y^{\rm res}_{\overline{\nu}_\tau} \lesssim  y_{\rm  peak}$, 
at least for sufficiently small $s^2_\tau$ (as illustrated in an example
in fig.~\ref{fig:evol1}).

The expressions  for the derivatives   of the lepton  asymmetries can be 
approximated, keeping only the dominant terms, in the following form:
\begin {equation}
 {d L_{\nu_\tau} \over dt } \simeq  
-{1 \over 4\,\zeta(3)}~\int dy\;y^2~f_{eq}^0 (y) ~
[\Gamma_{\tau s} - \overline {\Gamma}_{\tau s} ]
\end {equation}
\begin {equation}
 {d L_{\nu_\mu} \over dt } \simeq  
-{1 \over 4\,\zeta(3)}~\int dy\;y^2~f_{eq}^0 (y) ~
[\Gamma_{\mu s} - \overline {\Gamma}_{\mu s} ]
\end {equation}

Neglecting furthermore the oscillations of  $\overline{\nu}_\mu$'s  that are 
non resonant  and the  oscillations
of the $\nu_\tau$'s  
that  are  happening  at large  values    of $y$,
and  calculating the integrals  with the saddle  point approximation
since they are  dominated by the resonance  contributions, 
equation  (\ref{eq:limit2})      can  be rewritten as:
\begin {eqnarray}
 & ~& -2 \,\left [ \left ( 1 +  {(G_F^2 k_\mu)^2 \,T^{12} \,y^4_{\rm peak}
\over |\delta m^2_\mu|^2}
 \right)^ {1\over 2}
~ \left ( y_{\rm peak} \,  {b_\mu \, T^6 \over
 |\delta m^2_\mu| } \right ) \right]^{-1} 
\nonumber \\
&~&   ~~~~~~~
+ s^2_\tau \left [ \left ( 
s^2_\tau +  { (G_F^2 k_\tau)^2 \, T^{12} \,y^4_{\rm peak}
 \over |\delta m^2_\tau|^2} \right)^{1\over 2}
~ \left ( y_{\rm peak}  \left ( 2 + { L^{(\tau)} \over L^{(\mu)}} \right ) \,
 {b_\tau T^6 \over |\delta m^2_\tau| } \right ) \right]^{-1}  = 0
\end{eqnarray}

For $(G_F^2 k_\tau)^2 \, T^{12}\,y^4_{\rm peak}/ |\delta m^2_\tau|^2 \gg s^2$
and 
$(G_F^2 k_\mu)^2 \, T^{12} \,y^4_{\rm peak}/ |\delta m^2_\mu|^2 \gg 1$
the dependence on   $y_{\rm  peak}$  drops away
(showing that the choice of that  particular value  was not
critical)  and the condition for the  boundary 
takes the simple  form
\begin {equation}
s^2_\tau \, |\delta m^2_\tau|^2 \simeq  2 \, \left ( 2 
+ {L^{(\tau)} \over L^{(\mu)}} \right ) |\delta m^2_\mu |^2 \;.
\end{equation}

In this  way we have   reproduced  the functional  
form
($s^2_\tau \, |\delta m^2_\tau|^2 \simeq  {\rm const}~|\delta m^2_\mu |^2$)
of the boundary between  the regions  in parameter  space where the 
$\nu_\mu \leftrightarrow \nu_s$ oscillations  develop
or not  (see fig.~\ref{fig:final}).
The  numerical    results  can  be fitted 
with   the constant approximately  equal to $28$, and 
with  the  analytic  estimate  sketched  above 
we  obtain   a  reasonable approximation for it.
In fact, neglecting the small term 
$L_{\nu_e} - B_p/2$, one has:
$L^{(\tau)} = 2 L_{\nu_\tau} + L_{\nu_\mu}$
and 
$L^{(\mu)} = 2 L_{\nu_\mu} + L_{\nu_\tau}$;
since  $L_{\nu_\mu}$ is   negative 
we  have
$L^{(\tau)}/L^{(\mu)} \ge 2$.

It is also possible to understand why the thick lines in
fig.~\ref{fig:final} are not straight for relatively large
$s^2$, and bend upwards. This is due to the fact that in this 
situation the resonant momentum $y^{\rm res}_{\overline{\nu}_\tau}$ 
at the critical temperature is larger than $y_{\rm peak}$, and therefore
the approximation of equal $y^{\rm res}_\mu$ and 
$y^{\rm res}_{\overline{\nu}_\tau}$ fails. The constraint becomes more 
restrictive because the tau--sterile antineutrino oscillations are less 
effective, and a larger $\delta m^2_\tau$ is needed to compensate this 
effect.

\subsection {Validity of the approximations}

It is appropriate at this point to comment again on the validity of
the approximation scheme used in this  work.
As  discussed in section \ref{sec:QKE}, 
the Pauli--Boltzmann  system  of
equations describes correctly the flavor evolution of the  neutrino
ensemble only if the effective 
mixing in matter is  negligibly small
or  (for  large  mixing) if 
$\ell_{int}/\ell_m \ll 1 $,   that is    if   
the interaction length is much  shorter that the oscillation length 
and  the oscillations  cannot develop  coherently.
We can now   check a posteriori
that these conditions are  indeed  well satisfied 
for  neutrino  oscillation parameters
in  the `allowed  region'    indicated  in fig.~12, 
during the  crucial  epoch  
when  the    development  of the
$\nu_\mu \leftrightarrow \nu_s$ oscillations   is decided;
therefore    the  use  of
the  approximation  is  valid,  and the determination of the allowed
region is  correct.

As  discussed   before    the crucial moment
in the   flavor evolution of the neutrino  populations
is  around a  temperature 
 $T^*_\tau \simeq c_\tau |\delta m_\tau^2|^{1\over 6}/b_\tau$
determined  by the  $\delta m^2_\tau$   mass  difference.
Around this  temperature  
${\nu}_\tau \to {\nu}_s$ and
$\overline {\nu}_\tau \to \overline{\nu}_s$
transitions begin to proceed  with  slightly different  rates, and a 
positive $L_{\nu_\tau}$ asymmetry    starts  to  be  generated.
If  the    transitions  $\nu_\mu \to \nu_s$  do  not  proceed
sufficiently rapidly   to overcompensate 
the  growth of  $L_{\nu_\tau}$  when it is   still very small,
a  large $L^{(\mu)} \simeq L_{\nu_\tau}$  develops   that 
supresses  the  effective  $\nu_\mu$--$\nu_s$ mixing
for all  lower  temperatures.
On the contrary, if  the $\nu_\mu \to \nu_s$ oscillations
are  sufficiently  effective 
to  reverse the growth  of   $L^{(\mu)}$, 
the medium does  not  develop a large  asymmetry, and
at lower  temperatures ($T \lesssim |\delta m_\mu^2|^{1\over 6}/b_\mu$)
the oscillations  can  proceed  without  being  suppressed
by matter  effects,  resulting in a large population of sterile
neutrinos  at the  nucleosynthesis  epoch.

We can now investigate the  validity of  our  framework for
$T \sim T^*_\tau$.
For  this  temperature,  the effective  mixing  in matter  is 
very  small  (also  in the case of the 
$\nu_\mu \leftrightarrow \nu_s$ oscillations
where  $s_\mu^2 = 1$), except  for   narrow  regions  around the
resonant  momenta.  Therefore to  check the validity
of  our  framework  we have to check  that 
$\ell_{\rm int}/\ell_m$ is  indeed  small  for the
neutrinos  at the resonances.
Using the definitions of $\ell_{\rm int}$ and
$\ell_m$   we can  write  explicitely this condition as:
\begin{equation}
\left ( {\ell_{int} \over \ell_m} \right )_{\rm res}  =
{|\delta m^2_\alpha|\, s_\alpha \over 2 \,G_F^2 \, k_\alpha \,T^6 \,y^2} 
\ll 1
\label{eq:lintlm}
\end{equation}

For the  $\nu_\tau \leftrightarrow \nu_s$  oscillations,
for  small   asymmetry of the  medium,
(as it is the case  for $T \simeq T^*_\tau$)
the resonant  condition is:  
$b_\tau \,T^6\, y^2/|\delta m^2_\tau| \simeq c_\tau$,
and  substituting in  (\ref{eq:lintlm})
one  obtains the  condition
\begin{equation}
{s_\tau \over c_\tau } \ll {2 \,G_F^2 \,k_\tau \over b_\tau}
\simeq  0.04
\end{equation}
In the allowed region  we have $s_\tau \lesssim 3 \times 10^{-3}$,
and this condition is  always  satisfied.

For the $\nu_\mu \leftrightarrow \nu_s$  oscillations
the  mixing in matter 
is  negligibly small for  all momenta  $y$  except
for    small regions around $y = 0$,  and  
in case of a non symmetric medium around 
a second  (non vanishing) value, (\ref{y_res}).
The first  region is  not significant  because
for  $T \sim T^*_\tau$, it is    narrow  
(the  mixing is large  only  for values 
$y \lesssim \sqrt { |\delta m_\mu^2 /\delta m_\tau^2|}$)
and  for  the  values of the mass differences  we are  considering it
contains a  neglibly small   population  of   neutrinos.  
For the second  region, substituting  
$T \simeq T^*_\tau$
the condition (\ref{eq:lintlm})
can be  rewritten as:
\begin {equation}
 {1 \over y^2}  ~\left | { \delta m^2_\mu \over
 \delta m^2_\tau} \right | \ll   0.04
\label{eq:limx}
\end{equation}
The boundary of the  allowed  region   is    approximately 
described  by the relation (\ref{eq:approx}),
substituting $|\delta m^2_\mu/\delta m^2_\tau |$ in (\ref{eq:limx}) 
one obtains
\begin {equation}
{s_\tau \over y^2 } \ll  0.21
\end{equation}
This condition  is  satisfied  (for  all  $y$  with  a   significant 
neutrino population) along most of the boundary line.
The Pauli--Boltzmann  approximation  begins  to  be  poor for $s_\tau$ large,
that is only at  the `lower'  corner  of the  allowed   region  where 
one can  expect small  deviations    from   the results obtained
solving the detailed   Quantum Kinetic Equations.

We  note  that 
if the   oscillation parameters  are outside the   allowed   regions,
and  the  medium   of the expanding universe remains  quasi--symmetric,
at sufficiently low  temperature the matter  effects  become   negligible
and the $\nu_\mu \leftrightarrow \nu_s$ oscillations  can  proceed
coherently with  maximal  mixing.
In this  situation the Pauli--Boltzmann approximation is
not  formally valid,   however in this  case  
the  final  result   is   equal populations of   sterile 
and  standard neutrinos,  independently  of  the method used.

The  validity of the  Pauli--Boltzmann 
approach is also very well confirmed  by the   calculation of 
the boundary of the allowed region performed
using the Quantum Kinetic Equations by Foot
 \cite{Foot99}, 
whose  results are in  good agreement with what we  
obtain here.  The only  small  deviations 
(the bound is slightly more stringent than ours)
are present in the lowest  part, 
where we  expect our  approximation to be less accurate.

We stress again that the advantage of simpler equations, allowing the 
use of analytical arguments to check the numerical results, is hard to 
overestimate in a complicated problem such as the one we are 
considering here.

\section{Discussion of the results and conclusions}

Our results are in good agreement with the numerical results
obtained in \cite{Foot97,Foot99}, except in the region of the highest
acceptable $s^2_\tau$.
The curves of \cite{Foot97},
calculated in a Pauli--Boltzmann approach, do not show an upper bending
as ours do, while the results obtained in \cite{Foot99} using QKE
are slightly more restrictive than our curves. 

\begin{figure} [t]
\centerline{\psfig{figure=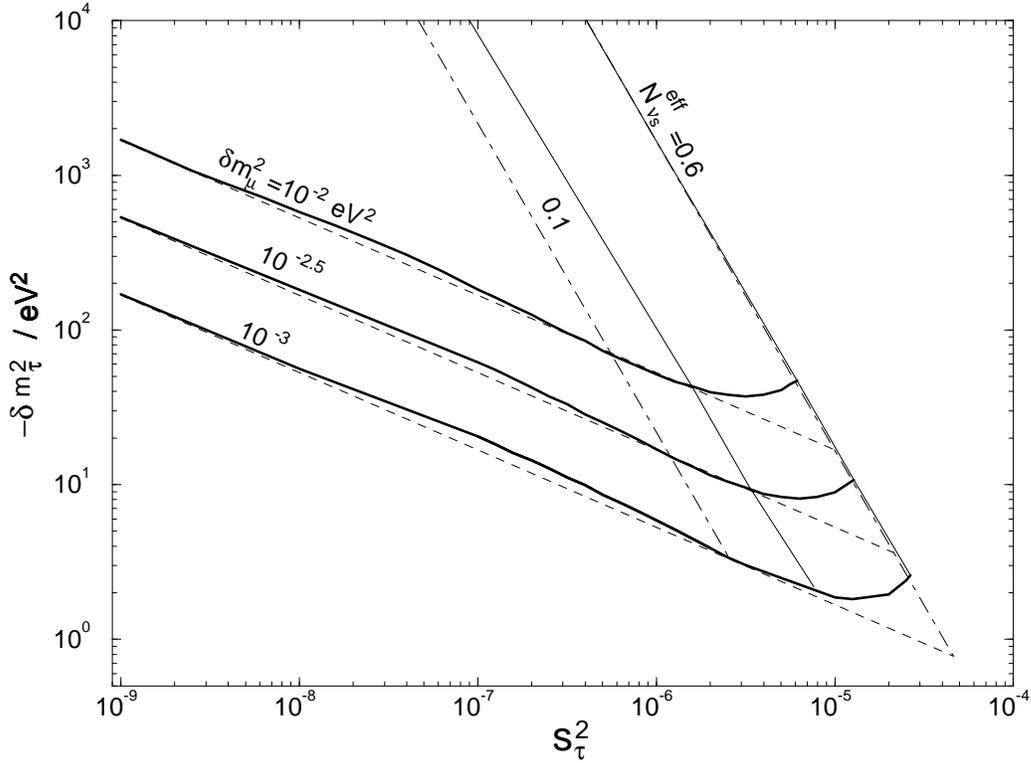,height=16cm,angle=270}}
\caption {Lines   corresponding  to  constant
$N_{\nu_s}^{\rm eff}$ in the  plane  ($s^2_\tau$, $\delta m^2_\tau$).
The  thin  solid  lines   correspond to 
$N_{\nu_s}^{\rm eff} = 0.1$   and 0.6; 
the  thin  dot--dashed lines   correspond to 
the  same  values of $N_{\nu_s}^{\rm eff}$ and have been calculated
in a  two neutrino  mixing   framework with negligible $L^{(\tau)}$.
The  three thick solid lines, calculated for the
three values 
$\delta m^2_\mu = 10^{-3}$, $10^{-2.5}$ and $10^{-2}$~eV$^2$, 
correspond to the  points  where the  
energy density of  sterile  neutrinos  passes quasi--discontinuosly
from the value  
$N_{\nu_s}^{\rm eff} = 1$   (below the curve)  to a   much
lower  value  (above the curve).  
The three dashed lines, corresponding to equation (\ref{eq:approx}), 
give a first approximation description of the numerical results.
\label{fig:final}
}
\end{figure}

\begin{figure} [t]
\centerline{\psfig{figure=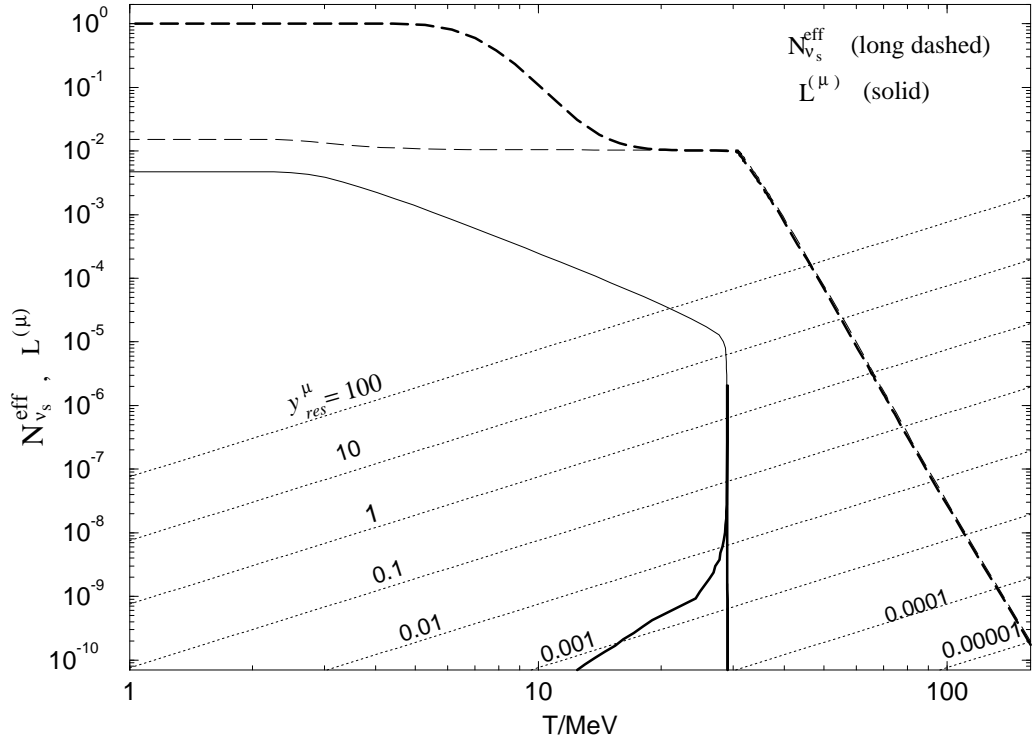,height=16cm,angle=270}}
\caption {Evolution the   asymmetry $L^{(\mu)}$  (solid  lines)
and the   sterile  neutrino   energy density  $N_{\nu_s}^{\rm eff}$
(dashed lines)  as a function of the  temperature.
The thin  curves are  calculated  for 
the set oscillation parameters
$\delta m^2_\mu = 10^{-2.5}$~eV$^2$, 
and  $s^2_\tau = 10^{-6}$, and 
$\delta m^2_\tau = 17.80$~eV$^2$, 
The thick   curves are  calculated  for 
the same  values  of 
$\delta m^2_\mu$  and
$s^2_\tau$   but with  a   slightly  smaller  
value $\delta m^2_\tau = 17.40$~eV.
\label{fig:boundary}
}
\end{figure}

\begin{figure} [t]
\centerline{\psfig{figure=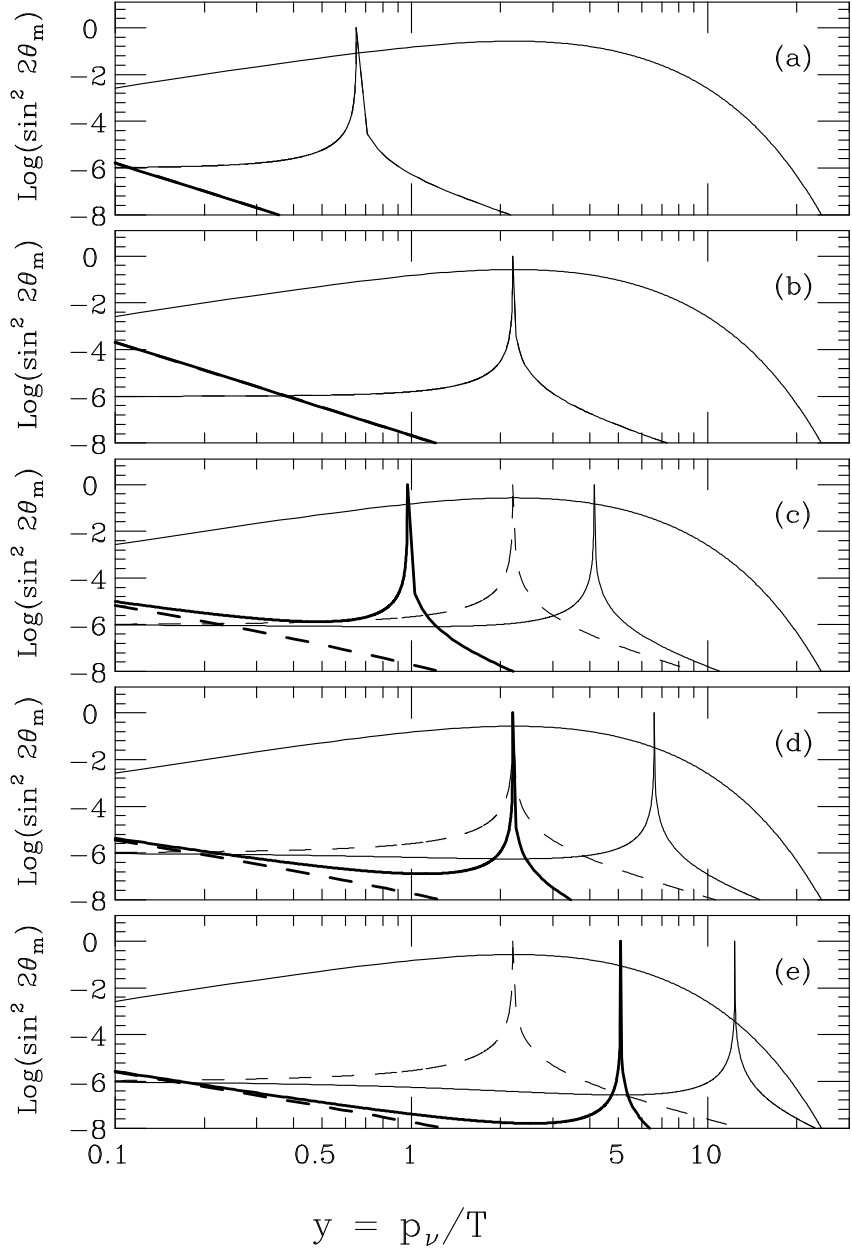,height=17.5cm}}
\caption {Mixing  parameters of  $\nu$'s and $\overline{\nu}$'s
in matter  for different
conditions  of the  medium.
The  oscillation   parameters
are: $\delta m^2_\mu = 10^{-3}$~eV$^2$, 
$s^2_\mu = 1$,
$\delta m^2_\tau = -10$~eV$^2$, 
$s^2_\tau = 10^{-6}$.
The  4  panels refer  to  the situations:
(a)  $T= 58.1$~MeV,  $L_{\nu_\tau} = 0$,
(b)  $T= 38.7$~MeV,  $L_{\nu_\tau} = 0$,
(c)  $T= 34.9$~MeV,  $L_{\nu_\tau} =  8.9 \times 10^{-7}$,
(d)  $T= 32.2$~MeV,  $L_{\nu_\tau} =  1.7 \times 10^{-6}$,
(e)  $T= 29.1$~MeV,  $L_{\nu_\tau} =  3.2 \times 10^{-6}$;
in all cases $L_{\nu_\mu} \simeq 0$.
In each panel  the  thick   (thin) curves    refer  to
$\nu_\mu \leftrightarrow \nu_s$   
($\nu_\tau \leftrightarrow \nu_s$)   oscillations,
the solid (dashed)  lines
to  neutrinos  (antineutrinos). The thin, smooth curve is a plot
of the function $y^2\,f_{eq}^0 (y)$.
\label{fig:film} }
\end{figure}

Our results do not agree 
with the semi-analytical calculations presented in \cite{Shi99a,Shi99b}, 
that obtain a much more stringent lower limit on the tau--neutrino mass 
($m_{\nu_{\tau}}\gtrsim 15$~{\rm eV}), that would give too much energy
density in the form of hot dark matter and problems for early 
structure formations.
The authors of \cite{Shi99a,Shi99b} claim that very small values of 
$y^{\rm res}_{\nu_\mu}$ are most important to  determine the fate of
the lepton number and that an insufficient resolution in the momenta 
may well miss this contribution and thus yield wrong results.
We have accurately tested our numerical results; for example,
the curves in figure~\ref{fig:boundary} describing the behaviour of
the asymmetry $L^{(\mu)}$ remain the same, if we change the minimal value 
of $y$ used to calculate the sterile neutrino distribution 
down to $10^{-6}$. This strengthens our  result, that
the values of $y^{\rm res}_{\nu_\mu }$ that can better destroy the 
lepton number are around $y_{\rm peak}$ and not the small ones. 
We performed the same test in several points in parameter space, that
essentially cover the thick curves in figure \ref{fig:final}. 
Moreover we note
that for small values of $y^{\rm res}_{\nu_\mu }$ (more precisely, for 
$y^{\rm res}_{\nu_\mu } < y^{\rm res}_{\overline{\nu}_\tau}
\sqrt{ |\delta m^2_\mu|/|\delta m^2_\tau|}
\simeq 10^{-3}-10^{-2}$ according to the region of parameters considered), 
the sterile neutrino production from $\nu_{\mu}-\nu_{s}$ transitions 
has a dominant non--resonant behaviour (and cannot therefore be evaluated 
in the resonant approximation), which is rather smooth and does not require 
a resolution particularly high to be estimated correctly. 

Furthermore, 
we do not agree with the condition imposed 
in \cite{Shi99a,Shi99b} to determine the range of parameters that allow the
survival of the lepton number generated. In those papers the amount 
of lepton number
destroyed by $\nu_{\mu} \leftrightarrow \nu_{s}$ oscillations for any 
$y^{\rm res}_{\nu_\mu }$ 
during the time needed to cross the resonance 
is compared with the total value of the lepton number $L^{(\mu)}$ 
produced up to that moment. 
In our opinion, the comparison should be done instead 
with the amount of lepton number
created by $\nu_{\tau} \leftrightarrow \nu_{s}$ oscillations during 
the same time interval (i.e. making a comparison of the two rates as we did). 
Actually this amount is comparable with the 
total lepton number created before and in fact it is even more, 
since the lepton number is exponentially growing. 

The same authors have also recently found another argument
that should restrict the allowed region \cite{Shi99c}. 
The QKE show a chaotic behaviour for the sign of the lepton number that is
generated at the critical temperature \cite{Shi96,Enq-chaos} and this would 
lead to the formation of lepton domains. Neutrino crossing the borders of 
these domains
would pass through a further resonance, with an additive contribution
to sterile neutrino production. We think that even though on
qualitative grounds this observation sets forth an important issue, from
a quantitative point of view only a full momentum dependent approach 
can give a definitive answer about the chaotic behaviour   
(moreover, the present results of momentum independent calculations 
\cite{Shi99c,Enq-chaos} are not in good agreement with each other). 
In \cite{Foot99} such an approach is used and the chaotic behaviour has also 
been found, but only for large enough mixing angles. 
A complete analysis
of the region in parameter space where chaos occurs is still missing however, 
and we think that the effect of lepton domain formation on the allowed 
region may be more accurately determined.

In conclusion, a momentum dependent Pauli-Boltzmann approach for 
active-sterile neutrino
oscillations in the Early Universe allows to derive, directly
from the equations, useful analytical results for the sterile neutrino 
production that support the numerical ones. An 
analytical procedure provides a deeper physical insight and allows a more
general picture of the possible solutions that sometimes pure numerical 
calculations hide (the generation of a relevant lepton number
missed in early numerical results is a good example). In this way we have 
been able to derive analytically in section \ref{sec:prod} 
the constraints from BBN on mixing parameters with the assumption of
negligible lepton numbers, showing that they essentially reproduce the 
results that can be obtained from QKE for $\nu_{\mu(\tau)}\leftrightarrow 
\nu_s$ oscillations.
When lepton numbers are present their effect is for most cases
a suppression of sterile neutrino production, but we showed that
it exists a regime where an enhancement is possible, even if 
in a parameter region not interesting for the present BBN bound.
 We were also able to give a full spectral description of the production, 
both in the case of negligible lepton numbers and when lepton numbers must be
considered. This allowed to specify better why in the case of a
two neutrino mixing the regions allowed by BBN do not change significantly 
with or without a generation of  lepton number. 
The most important result has been
obtained about the possibility that the generation of lepton number
through tau--sterile neutrino oscillations 
can suppress sterile neutrino production from mu--sterile
oscillations for the values of mixing parameters that could describe 
the Super--Kamiokande results. We confirmed numerically previous results
\cite{Foot97}, that show that this is possible with a lower limit on the
tau neutrino mass (of a few eV) compatible with present cosmological
observations. We could provide an analytical support to these results 
and show that one should not worry to obtain an exceedingly high resolution in
momentum space. This settles in our opinion the recent controversy on the 
minimum $\nu_\tau$ mass allowed. 

We did not consider the more general situation where
lepton domains can be created. We think that clear results about this
issue are necessary, but still missing. Anyway it is surely important to 
have established results in the simpler homogeneous case.

\vspace {0.9 cm}

\noindent {\bf Acknowledgments}  \\
We would like to thank
E.Kh. Akhmedov, R. Foot  and R.R. Volkas for useful discussions.

\newpage

\end{document}